\documentclass{JINST}

\usepackage{url}
\usepackage{textcomp}
\usepackage{wrapfig}
\usepackage{nth}

\graphicspath{ {./} {./pix/} }

\title{Design and Operation of FACT --\\ The First G-APD Cherenkov Telescope}

\newcommand{\ethz}{$^a$}
\newcommand{\tudo}{$^b$}
\newcommand{\unige}{$^c$}
\newcommand{\uniw}{$^e$}
\newcommand{\epfl}{$^d$}

\newcommand{\unigea}{$^{,c}$}
\newcommand{\epfla}{$^{,d}$}
\newcommand{\uniz}{$^{1}$}
\newcommand{\unizh}{$^{^1}$}
\newcommand{\kynu}{$^{2}$}
\newcommand{\kynuh}{$^{^2}$}
\newcommand{\mpim}{$^{3}$}
\newcommand{\mpimh}{$^{^3}$}
\newcommand{\tum}{$^{4}$}
\newcommand{\tumh}{$^{^4}$}

\newcommand{\aut}{$^*$}

\author{
H.~Anderhub\ethz,
M.~Backes\tudo,
A.~Biland\ethz,
V.~Boccone\unige,
I.~Braun\ethz,
T.~Bretz\ethz\epfla\thanks{Corresponding authors: {\tt thomas.bretz@phys.ethz.ch, qweitzel@phys.ethz.ch}} , 
J.~Bu\ss\tudo,
F.~Cadoux\unige,
V.~Commichau\ethz,
L.~Djambazov\ethz,
D.~Dorner\uniw\unigea,
S.~Einecke\tudo,
D.~Eisenacher\uniw,
A.~Gendotti\ethz,
O.~Grimm\ethz,
H.~von Gunten\ethz,
C.~Haller\ethz,
D.~Hildebrand\ethz,
U.~Horisberger\ethz,
B.~Huber\ethz\unizh,
K.-S.~Kim\ethz\kynuh,
M.~L.~Knoetig\ethz,
J.-H.~K\"ohne\tudo,
T.~Kr\"ahenb\"uhl\ethz,
B.~Krumm\tudo,
M.~Lee\ethz\kynuh,
E.~Lorenz\ethz\mpimh,
W.~Lustermann\ethz,
E.~Lyard\unige,
K.~Mannheim\uniw,
M.~Meharga\unige,
K.~Meier\uniw,
T.~Montaruli\unige,
D.~Neise\tudo,
F.~Nessi-Tedaldi\ethz,
A.-K.~Overkemping\tudo,
A.~Paravac\uniw,
F.~Pauss\ethz,
D.~Renker\ethz\tumh,
W.~Rhode\tudo,
M.~Ribordy\epfl,
U.~R\"oser\ethz,
J.-P.~Stucki\ethz,
J.~Schneider\ethz,
T.~Steinbring\uniw,
F.~Temme\tudo,
J.~Thaele\tudo,
S.~Tobler\ethz,
G.~Viertel\ethz,
P.~Vogler\ethz,
R.~Walter\unige,
K.~Warda\tudo,
Q.~Weitzel\ethz\aut,
M.~Z\"anglein\uniw\\
\\
\llap{\ethz}ETH Zurich, Switzerland~~\,--\,~~Institute for Particle Physics, Schafmattstr.~20, 8093 Zurich\\
\llap{\tudo}Technische Universit\"at Dortmund, Germany \\
Experimental Physics 5, Otto-Hahn-Str.~4, 44221 Dortmund\\
\llap{\unige}University of Geneva, Switzerland \\
ISDC Data Center for Astrophysics, Chemin d'Ecogia~16, 1290 Versoix\\ D\'epartement de Physique Nucl\'eaire et Corpusculaire, Quai Ernest-Ansermet 24, 1211 Geneva\\
\llap{\epfl}EPF Lausanne, Switzerland~~\,--\,~~Laboratory for High Energy Physics, 1015 Lausanne\\
\llap{\uniw}Universit\"at W\"urzburg, Germany \\
Institute for Theoretical Physics and Astrophysics, Emil-Fischer-Str.~31, 97074 W\"urzburg,\\
{\tiny
\llap{}{}\\
\llap{\uniz}{Also at: University of Zurich, Physik-Institut, 8057 Zurich, Switzerland}\\
\llap{\kynu}{Also at: Kyungpook National University, Center for High Energy Physics, 702-701 Daegu, Korea}\\
\llap{\mpim}{Also at: Max-Planck-Institut f\"ur Physik, 80805 Munich, Germany}\\
\llap{\tum}{Also at: Technische Universit\"at M\"unchen, 85748 Garching, Germany}\\
}
}

\abstract{The First G-APD Cherenkov Telescope (FACT) is designed
to detect cosmic gamma-rays with energies from several hundred GeV up
to about 10\,TeV using the Imaging Atmospheric Cherenkov
Technique. In contrast to former or existing telescopes, the camera of
the FACT telescope is comprised of solid-state Geiger-mode Avalanche
Photodiodes (G-APD) instead of photomultiplier tubes for photo
detection. It is the first full-scale device of its kind employing
this new technology. The telescope is operated at the Observatorio del Roque de
los Muchachos (La Palma, Canary Islands, Spain) since fall 2011. This
paper describes in detail the design, construction and operation of
the system, including hardware and software aspects. Technical
experiences gained after one year of operation are
discussed and conclusions with regard to future projects are drawn.
}


\begin{document}

\newpage
\section{Introduction}\label{sec:intro}

For more than two decades, Imaging Atmospheric Cherenkov Telescopes
(IACT) have been very successful in observing cosmic gamma-ray sources
at very high energies (VHE, about 50\,GeV to 100\,TeV)
\cite{lorenz12}. More than 100 sources have been detected, including
galactic, as well as extragalactic, objects like Supernova Remnants or
Active Galactic Nuclei (AGN). The observed gamma-ray flux is energy
and source dependent and can be highly variable.
Combining the datasets with observations at other wavelengths
(e.g.\ radio or X-ray data) allows to study cosmic particle
acceleration and radiation emission mechanisms. To detect variability
on long time scales and understand the flaring behavior of AGN,
continuous monitoring for several months or years is necessary. This can
be achieved by robust and small telescopes \cite{bretz08},
complementary to the large-scale arrays serving many tasks.

Cosmic-ray induced particle cascades generate Cherenkov light in the
atmosphere \cite{weekes89} with a broad spectrum peaking between
300\,nm and 350\,nm. These light flashes have a typical
photon density at ground of a few up to several hundred photons per
square-meter on nano-second time scales. If recorded with an imaging device,
the image topology
allows to reconstruct energy and direction of the primary particle and
to determine its type \cite{hillas85}. The latter allows the
suppression of the strong hadron induced background exceeding the rate
of gamma induced air-showers typically by $10^4$ to $10^5$.

In order to achieve the goal of long-term monitoring as good as possible,
operation during moonlit nights is mandatory. While for dark nights
the intensity of the diffuse night-sky background is of the order of
10\,MHz/cm$^2$ \cite{mirzoyan94,king}, for moonlit nights it can be three
or four orders of magnitude higher.

The Cherenkov light is collected by means of a large reflector with an
area of several to a few hundred square meters, focusing it onto a fast and
sensitive camera. To keep the costs and the point-spread function low,
reflectors are segmented. Due to the small opening angle of Cherenkov
light, the reflector has to be oriented towards the shower origin. 
This can be achieved, for example, by an alt-azimuth mount able to track any position
on the sky.

Up to now, all cameras of IACTs have been built
with photomultiplier tubes (PMT), but during the last
years novel solid-state photosensors have been developed
\cite{renker09}. Primarily, Geiger-mode Avalanche Photodiodes (G-APD, notation
used in this paper for entire sensor subdivided in cells)
have the potential to replace PMTs in gamma-ray astronomy. They offer
a high gain ($10^5$ to $10^6$) and can be operated under much brighter
light conditions allowing observations during moon time. Their single
photon counting capability is a powerful tool for the signal
calibration, and their compactness and their low operational voltage ($<$\,100\,V)
simplify the camera design. The photon detection efficiency (PDE) of
commercially available G-APDs is already at the level of the best
PMTs, and is still improved by the manufacturers.

In order to demonstrate the advantages of novel light sensors for VHE
gamma-ray astronomy, the First G-APD Cherenkov Telescope (FACT) has
been developed featuring a complete camera based on G-APDs. After
successful tests with a small (36 pixels) prototype module in summer
and fall 2009 \cite{anderhub09} and a re-design, the full-size camera
(1440 pixels) has been constructed. In fall 2011 the device has
been shipped to La Palma (Canary Islands, Spain), and installed on an
existing telescope mount at the Observatorio del Roque de los
Muchachos (ORM, 2200\,m a.s.l.). In October 2011, few hours after installation,
first air-shower images have been recorded. The system has been
running almost every night since then and several source observations
have already been conducted for its commissioning.

\section{System Overview}\label{sec:overview}

\begin{figure}[tb]
    \centering
    \includegraphics[width=\textwidth]{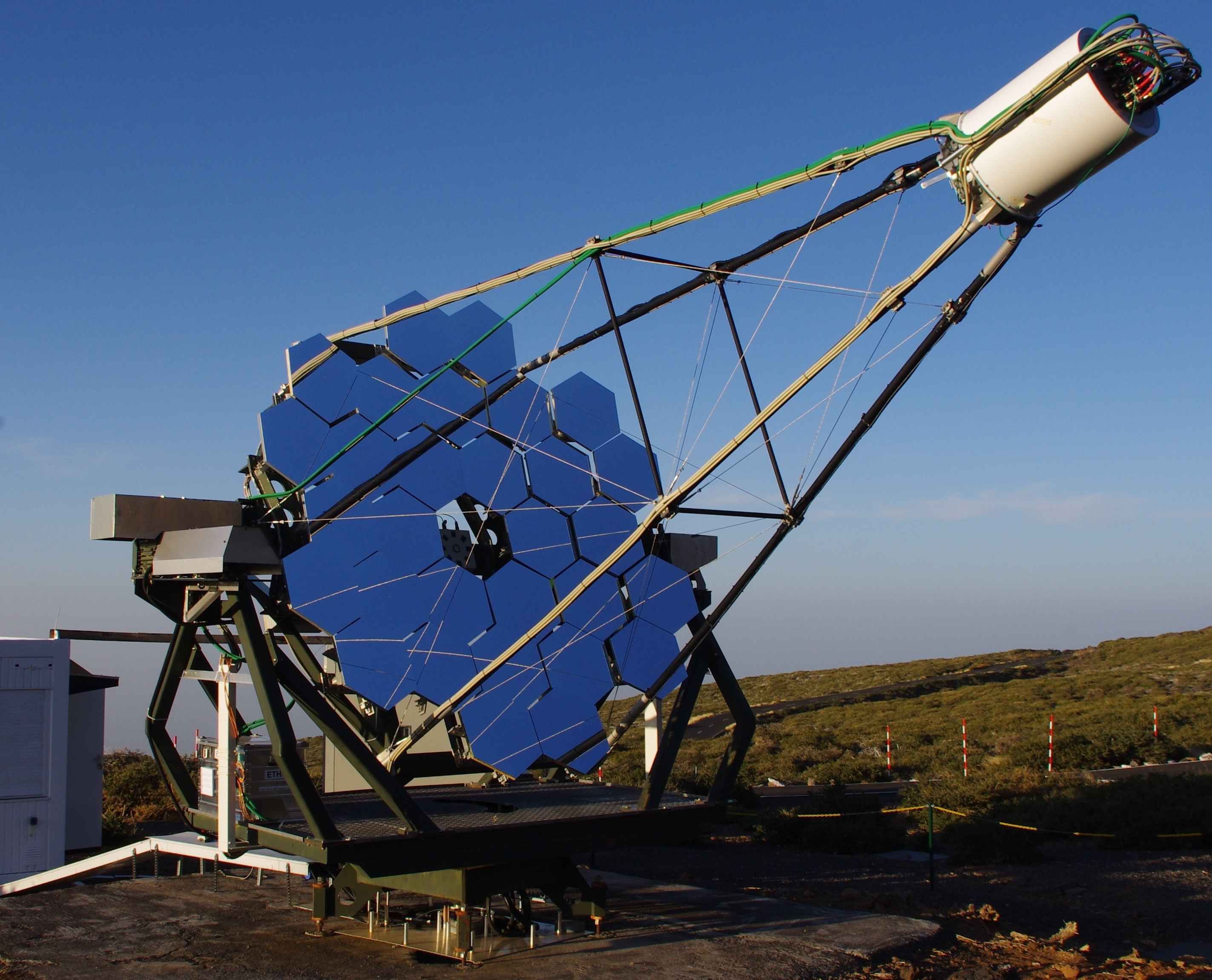}
    \caption{Photograph of the FACT telescope at the
      Observatorio del Roque de los Muchachos on La Palma
      (Canary Islands, Spain). At
      the top right the camera is visible (white cylinder), including
      the cables for power and communication connected at its
      back. The cables are fixed along the telescope masts and,
      together with the cables for the drive system, guided through a
      cable duct to the counting house (the container at the
      left). In the middle of the reflector where one mirror is
      missing, the light pulser and the video-camera are
      installed. The cooling unit is inside the metal box placed on
      the telescope platform.}
    \label{fig:FACT_telescope}
\end{figure}

\begin{figure}[b]
    \centering
    \includegraphics[width=\textwidth]{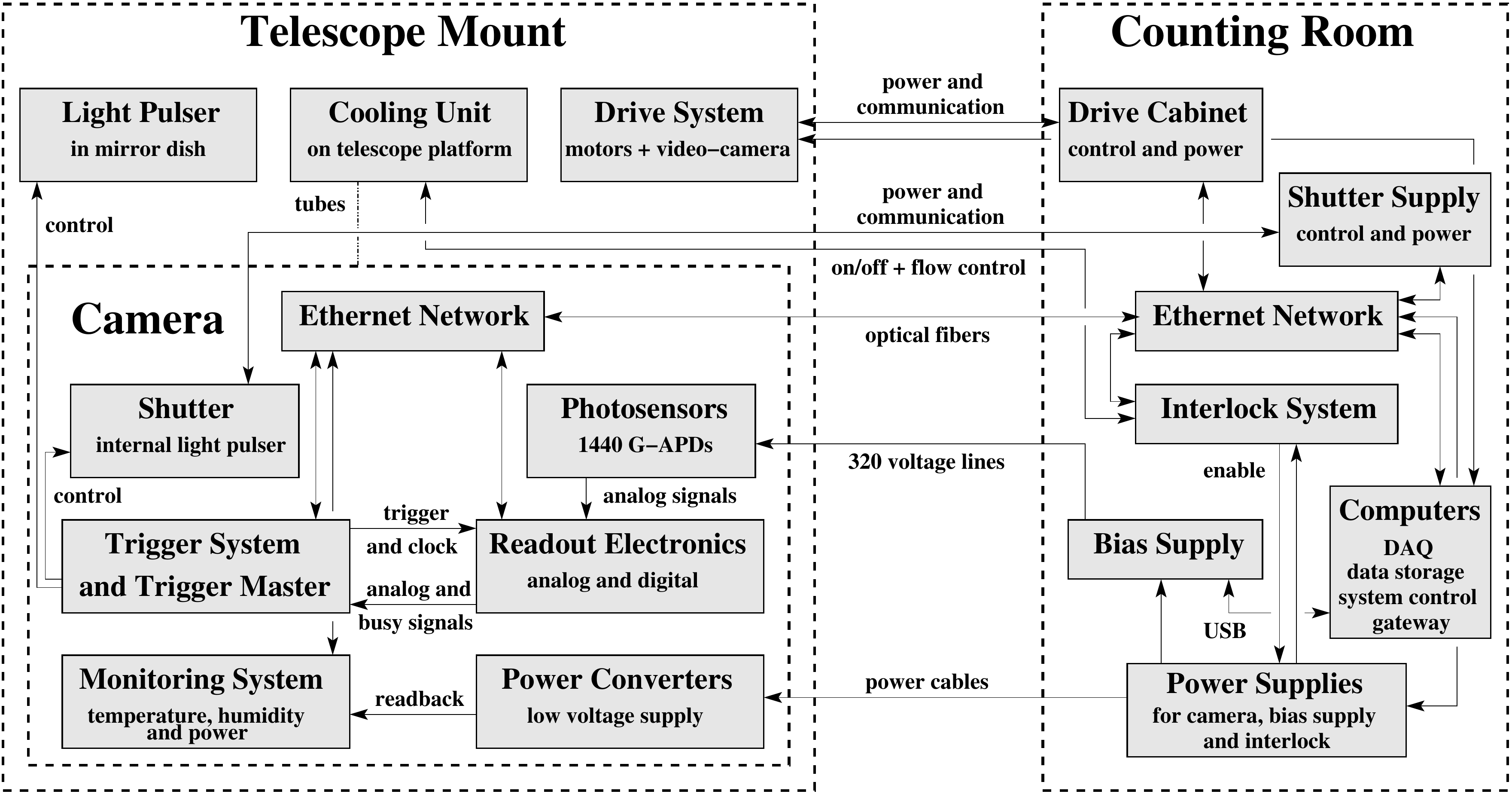}
    \caption{Schematic overview of the different sub-systems (gray boxes)
      constituting the FACT telescope. The arrows indicate electrical
      connections for power, signal transfer or communication. All components
      are located either in the counting room (right) or on the telescope
      mount (left). Hardware belonging to the camera is represented as a
      subgroup of the telescope mount.}
    \label{fig:system_overview}
\end{figure}

The FACT experiment consists of several sub-systems serving different
purposes during operation. In this section an overview of the setup is
given with links to detailed descriptions following in the next
sections. All hardware components are either located on the telescope
mount or in a counting room a few meters away. Figure~\ref{fig:FACT_telescope}
shows a photograph of the telescope. In figure~\ref{fig:system_overview} a schematic diagram of the system is presented. The installations at the ORM are located 
on the site
of the Major Atmospheric Gamma-ray Imaging Cherenkov Telescopes
(MAGIC, \cite{magic}), from which power is
provided.
The Davies-Cotton reflector and the camera are installed on a movable
mount. A fast light pulser for calibration purposes is situated in the
mirror dish (see section~\ref{sec:telescope}). A video-camera, installed
next to the light pulser, allows to take pictures of the
telescope-camera to check the environment conditions and to calibrate
the pointing using the reflection of bright stars (c.f.\
section~\ref{sec:telescope:calibration}).

The camera has a total field-of-view of 4.5\textdegree{} and is
comprised of 1440 individual pixels equipped with a G-APD and a
light concentrator each (see
section~\ref{sec:camera}). All channels are read out individually. The
required electronics for analog processing, digitization, and
triggering is integrated into the camera body. Trigger signals, as
well as a precise clock signal, are distributed from the trigger master
board to 40 digitizer boards (see
section~\ref{sec:camera:electronics}). The environmental conditions inside
the camera, such as temperature and humidity, and all supplied
voltages and drawn currents from the internal power-supplies are
monitored by a slow control board. Communication and data readout is
based on standard Ethernet connections. Waste heat, dissipated by the
electronics and internal power supplies, is conducted through a
water-circulation system (see section~\ref{sec:camera:aux}). Its pump is
placed on the telescope platform to limit the length of the connecting
tubes.
The operation voltage of the photo sensors, the so-called bias
voltage, is supplied to groups of four and five sensors and provided
by a dedicated power supply located in the counting room. During data
taking this voltage can be regularly adjusted in order to keep the gain of the
sensors stable (see section~\ref{sec:camera:aux}). The counting room also
hosts the power supplies for the camera and the drive motors, as well
as the control system of the latter. An interlock system ensures that
the camera is only powered when the cooling unit is operating. In
order to allow remote operations, both the power unit of the camera
and the interlock system can be accessed via Ethernet. Several
computers are employed for data acquisition, temporary data storage
and system control via online software programs (see
section~\ref{sec:software}). A graphical user interface is provided for
the telescope operator. Each day the raw data from the last night is
automatically transferred from the telescope site to a data center in
mainland Europe, where it is processed and prepared for the offline
analysis. In order to obtain preliminary results quickly, an automated analysis
is running on the on-site computers.

\section{Telescope}\label{sec:telescope}

\subsection{Mount and Drive}\label{sec:telescope:drive}

As a mount, the former HEGRA CT\,3 telescope,
c.f.\ \cite{Hegra1,Hegra2}, is used. The alt-azimuth mount is made of a
massive steel structure driven by the originally installed industrial
worm gears (ZAE-Antriebssysteme GmbH \& Co KG, i=46).
 All other components were upgraded to a drive system
similar to the one in use for both MAGIC telescopes
\cite{Bretz2009}. Due to the more than ten times lower telescope weight (compared to MAGIC), and thus
reduced requirements on the motor power, 1\,kW motors (Bosch,
MSK\,070-D-0300) with a maximum torque of 8\,Nm are employed.  Their
power is transmitted by planetary gears (Wittenstein alpha,
LP-120-M02-30-111, i=30) to the worm gears connected on both axes.
Planetary gears and worm gears are connected by a transmission shaft
with a bevel gear allowing movements in case of failure with a
standard curtain crank. As motor controllers, Bosch IndraDrive HCS\,02
are used. They are controlled from a Programmable Logic Controller (PLC,
Bosch SPS L\,40) via Profibus connection. High-level commands are transmitted via
Ethernet from a standard PC. The position control loop is closed by
shaft-encoders, one on each axis (Heidenhain ROQ\,425 EnDat\,01). The
PLC program and the PC program are almost identical to the MAGIC
version. The main differences are due to different interlock systems.
A simple joystick, connected via cable to the PLC, is
available to move the telescope without PC. All components are watertight and
classified as IP\,67 (housing) or IP\,64 (shafts). An additional
splash water protection is installed.


\subsection{Pointing Calibration}\label{sec:telescope:calibration}

To correct for the imprecision of a real telescope mount, a pointing
calibration is performed. The resulting pointing model allows to
convert calculated source positions to real command values. For its
determination, the telescope is pointed to a sequence of many bright
stars, and the position of each reflected image on the camera surface
is evaluated w.r.t.\ their nominal position in the camera center. To
determine the camera center, eight Light Emitting Diodes (LED) are
mounted in a circle around the camera center at the camera edge.
Covering the sky homogeneously with such calibration measurements the
pointing correction can be parametrized depending on the nominal
pointing position. Once a pointing model has been computed and applied, the
calibration measurements
can also be used to evaluate the
current pointing quality (the residuals).

Before the start of operation, a first pointing model has been
determined. Measurements after almost one year of
operation have shown that the pointing accuracy during tracking has only
marginally degraded and is still better than 30\,arcsec.

\subsection{Reflector}

The existing reflector has the sphere-like shape of a so-called
Davies-Cotton mirror arrangement \cite{DaviesCotton} in which
individual mirrors are placed at their focal distance to the focal
point and oriented towards a point at twice the focal length. This
combines the advantages of the correct focal distance with an ideal
alignment, and is cheaper than a single-reflector design. The former glass
mirrors have been exchanged with the mirrors originally built for an
upgrade of HEGRA CT\,1.
The new mirrors are entirely made out of aluminum, with a honeycomb
inlay between the front and the back plates. They are of hexagonal
shape, covering an area of 0.317\,m$^2$ each. Being comprised of 30 of
such mirrors, the total reflective surface of the telescope amounts to
9.51\,m$^2$ (about 10\% more than for HEGRA CT\,3).

\subsubsection{Re-workings and Spectral Reflectivities}

\begin{figure}[htb]
    \centering
    \includegraphics[width=0.82\textwidth]{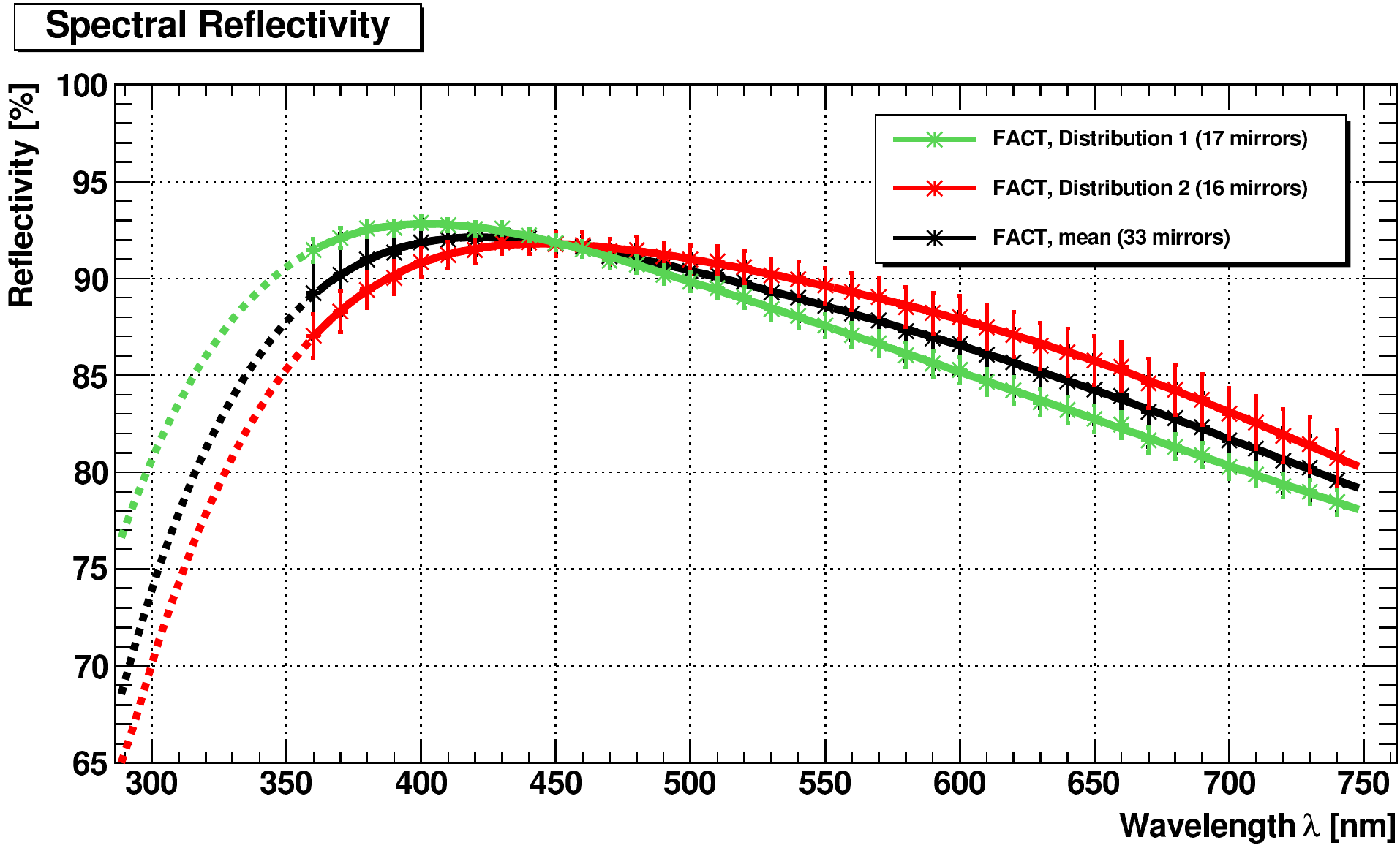}
    \caption{Spectral reflectivity for all 33 measured mirrors. The
      two distributions (green, red) originate from batches which were
      coated independently. The black curve is the average of all
      mirrors. Error bars represent the measured spread. The
      interpolation (solid) and extrapolation (dotted) was done by
      fitting a \nth{6} order polynomial.}
    \label{fig:refl}
\end{figure}

The more than twelve year old mirrors were re-machined using
diamond-milling by the company LT Ultra Precision Technology GmbH.
Subsequently, they were coated with SiO$_2$ at the
Fraunhofer Institute for Manufacturing Technology and Applied
Materials Research. Within a methane atmosphere of a
few mbar, silicon was deposited to the mirrors with a sputtering
technique and afterwards oxidized. Thus, SiO$_2$ was built up with some
admixture of carbon from the dissociation of methane. The coating
thickness has a major influence on the spectral reflectivity by thin
layer interference and should be less than 120\,nm, taking the
spectral shape of the Cherenkov radiation into account. The specular
reflectivity of all mirrors was measured to be constant within 4\%
over the surface of every single mirror. The coating of the mirrors
was conducted in two batches affecting the coating thickness and
thus the spectral reflectivity. The different charges form two classes
of mean reflectivity as depicted in figure~\ref{fig:refl}.

\begin{figure}[htb]
    \centering
    \includegraphics[width=0.85\textwidth]{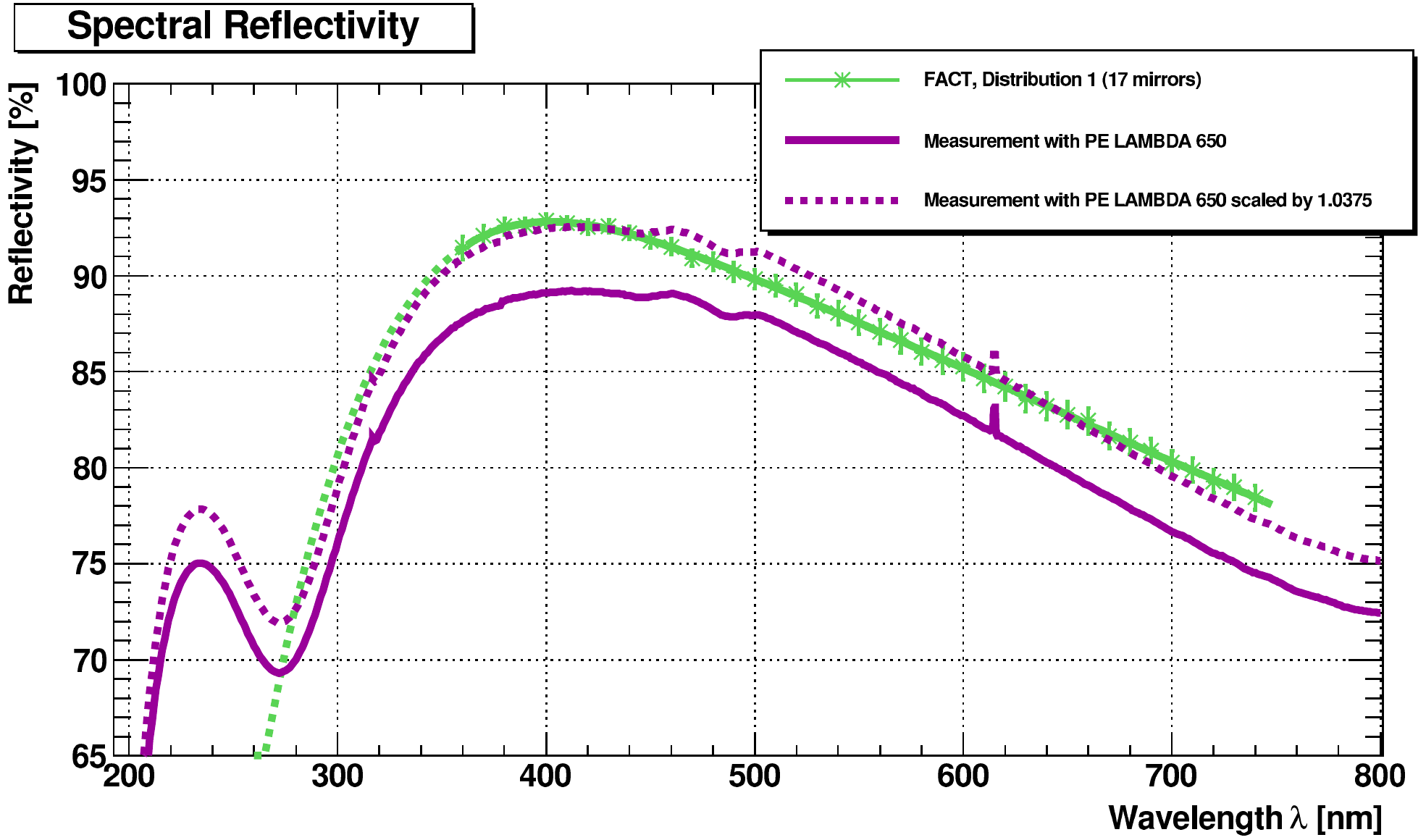}
    \caption{Spectral reflectivity for the first batch of mirrors
      (green) compared to the measurement of a single mirror using a
      different setup (down to 200\,nm, solid purple). An instrumental
      offset between both measurements is visible, well described by
      applying a scale factor (dashed purple).}
    \label{fig:refl_min}
\end{figure}

In order to determine the coating thickness, the reflectivity of a
single mirror has been remeasured with a PerkinElmer LAMBDA~650 UV/Vis
Spectrophotometer, which offers a wider wavelength range
of~190\,nm\,--\,900\,nm compared to 360\,nm\,--\,740\,nm of the former
measurements. The wavelength accuracy is $\pm$0.15\,nm. The measurements have
been conducted in steps of 1\,nm between 200\,nm and 800\,nm. These
results and the average of the corresponding batch measured previously
are shown in figure~\ref{fig:refl_min}. Clearly an inter-instrument
offset can be seen, probably due to slightly different integration
angles for the specular reflections. In order to cross-check the shape
of the extension of the batch mirror measurements to lower wavelengths
(green dashed line, used for simulations) a scaling factor of 1.0375
has been applied to the PerkinElmer measurement.

The interference minimum measured occurred at 272\,nm. Assuming an
incidence angle of~30\textdegree{}, the refractive index of SiO$_{2}$
at~272\,nm wavelength is reported as~1.589 for crystalline silica
and~1.50 for fused silica \cite{refractive}. Hence, the coating
thickness~$d$ is estimated as~90.2\,nm and~96.4\,nm.

\subsubsection{Focal Lengths and Point Spread Functions}

After the milling process some tension might persist inside the
aluminum plate and cause distortions to the spherical symmetry of the
mirror resulting in two different focal lengths for two orthogonal
directions. In order to measure the distance and size of the spot with the
smallest distortion, a test setup was built. As a point source, a red
LED was placed behind a 1\,mm pinhole at twice the expected focal
distance. Pictures of the image were taken from the front side of a
backside illuminated screen at several distances with a commercial
Sony~$\alpha$550 camera with an image resolution of
4,592$\times$3,056\,pixels and equipped with a SIGMA
MAKRO\,105\,mm\,F2.8\,EX\,DG objective lens.

For each background subtracted and noise suppressed image, the
smallest ellipse containing 95\% of the light is fitted. Plotting the
area of the ellipse versus distance allows to find the point of the
smallest distortion. In this way, the ideal distance of the mirror to
the focal plane, called focal length in the following, and the size of
the spot containing 95\% of the reflected light, in the following
called point-spread function (PSF) is defined.

\begin{figure}[htb]
    \centering
    \includegraphics[width=0.49\textwidth]{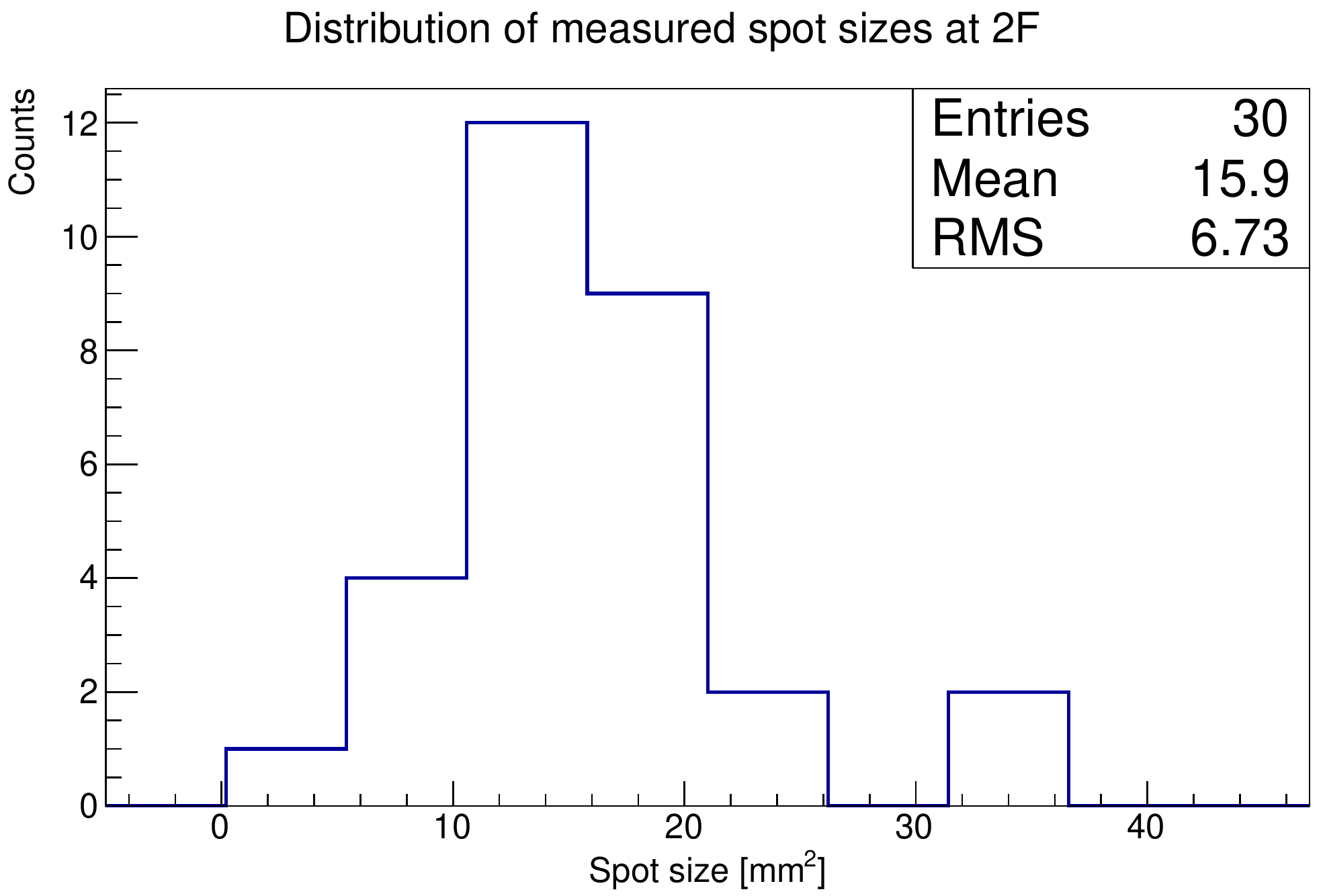}
    \hfill
    \includegraphics[width=0.49\textwidth]{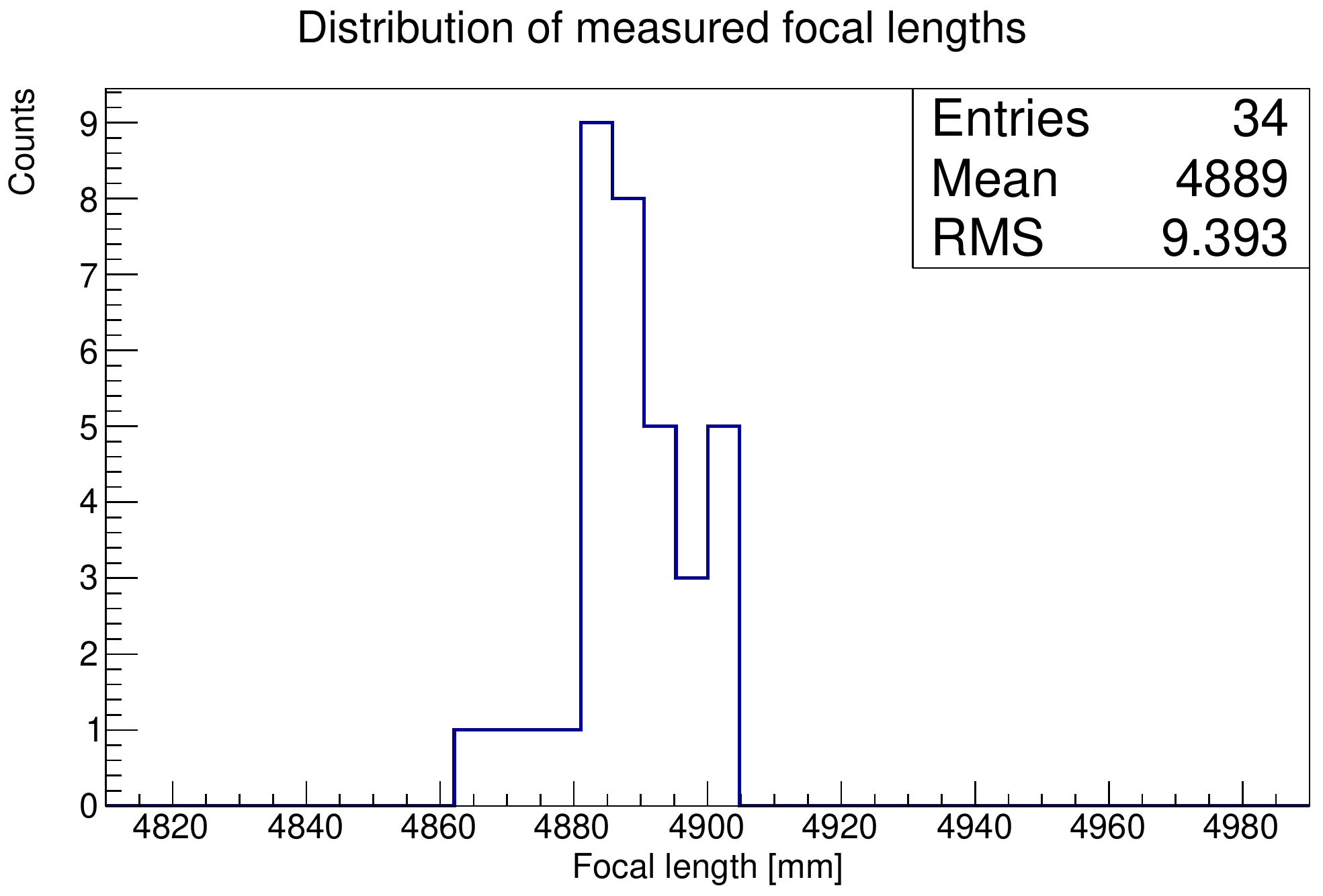}
    \caption{Left: Distribution of the area in square-millimeter of an
     ellipse containing 95\% of the light at 2$F$ for the 30 mirror facets.
     The corresponding value at their measured focal length is thus four
      times smaller.
      The focal length was determined as the point at which the area gets
      minimal. Right: Corresponding distribution of the focal
      length (for 30+4 facets).}
    \label{fig:focal-length}
    \label{fig:psf}
\end{figure}

Figure~\ref{fig:focal-length} (right) shows the resulting distribution
of focal lengths depicting a very small spread of 9\,mm ($<
2$\textperthousand) around the average focal length of
4.889\,m. Figure~\ref{fig:psf} (left) shows the distribution of the
measured point-spread functions.
The mean of the distribution of spot sizes measured at 2$F$ is
(15.95\,$\pm$\,6.73)\,mm$^2$. The spot size at their focal distance
is four times smaller. That means that nearly all of the mirror facets
focus 95\% of the reflected light in an area well below one quarter
of the size of one pixel (19.54\,mm$^2$).

\subsubsection{Alignment}

In order to be able to align all mirrors individually, each mirror is equipped
with a tripod which allows to alter the $z$-position at three
points. For the on-site alignment of the mirrors, an equipment has
been set up at a distance of twice the desired focal length along the
optical axis of the telescope, when the camera was not yet mounted.
To define the optical axis, a laser was shot from the center of
the reflector through the center of the camera mounting ring. The
alignment device consisted of a laser and a screen, both installed at
the same distance to the optical axis. A universal joint allowed to
orient the device towards each individual mirror. For a better
sensitivity the screen was monitored with a CCD camera also displaying
a cross hair for better alignment. With this method a typical
alignment error of the center of the light distribution of
$\pm$1.5\,mm on the focal plane was achieved. Admittedly, the
point-spread function is currently compiled from two mirror batches
about 4.5\,mm apart on the focal plane. These two batches correspond
to the upper and lower half of the mirrors and result from the bending
of the camera holding structure which was used to determine the
optical axis while the reflector was turned upside-down for an easier
access to the upper half of the mirrors. A new device allowing
alignment also with the camera mounted is under construction.

\subsection{Light-pulser}\label{sec:telescope:LP}

For test and calibration purposes the telescope uses two
light-pulser systems. In this section the so-called external
light-pulser is described, situated in the center of the mirror
dish. A second system, the internal light-pulser, is installed on the
inside of of the camera shutter (see
section~\ref{sec:camera:layout}). Both light pulsers are operated with
similar electronics.

\begin{figure}[hb]
    \centering
    \includegraphics[height=8.2cm]{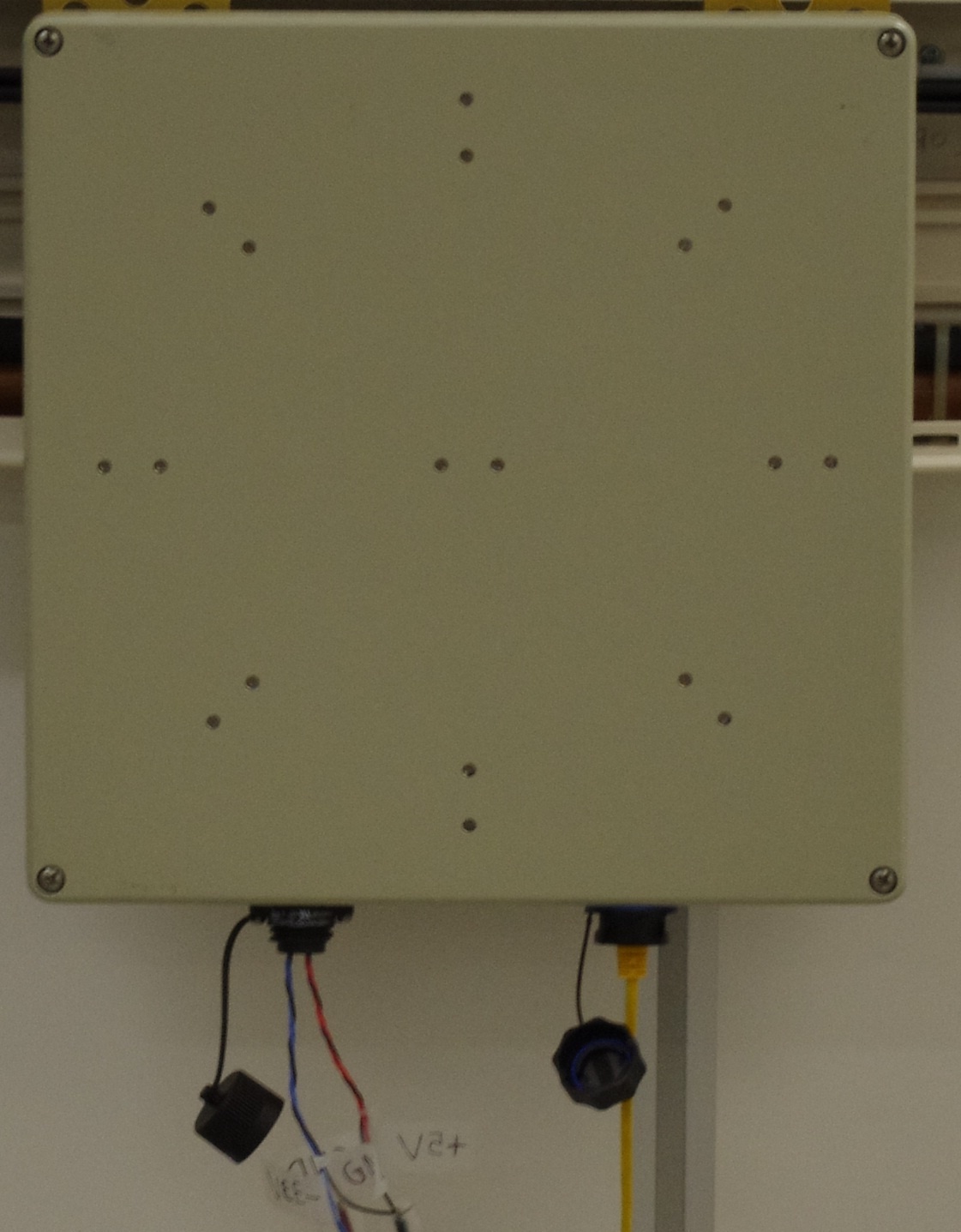}
    \hfill
    \includegraphics[height=8.2cm]{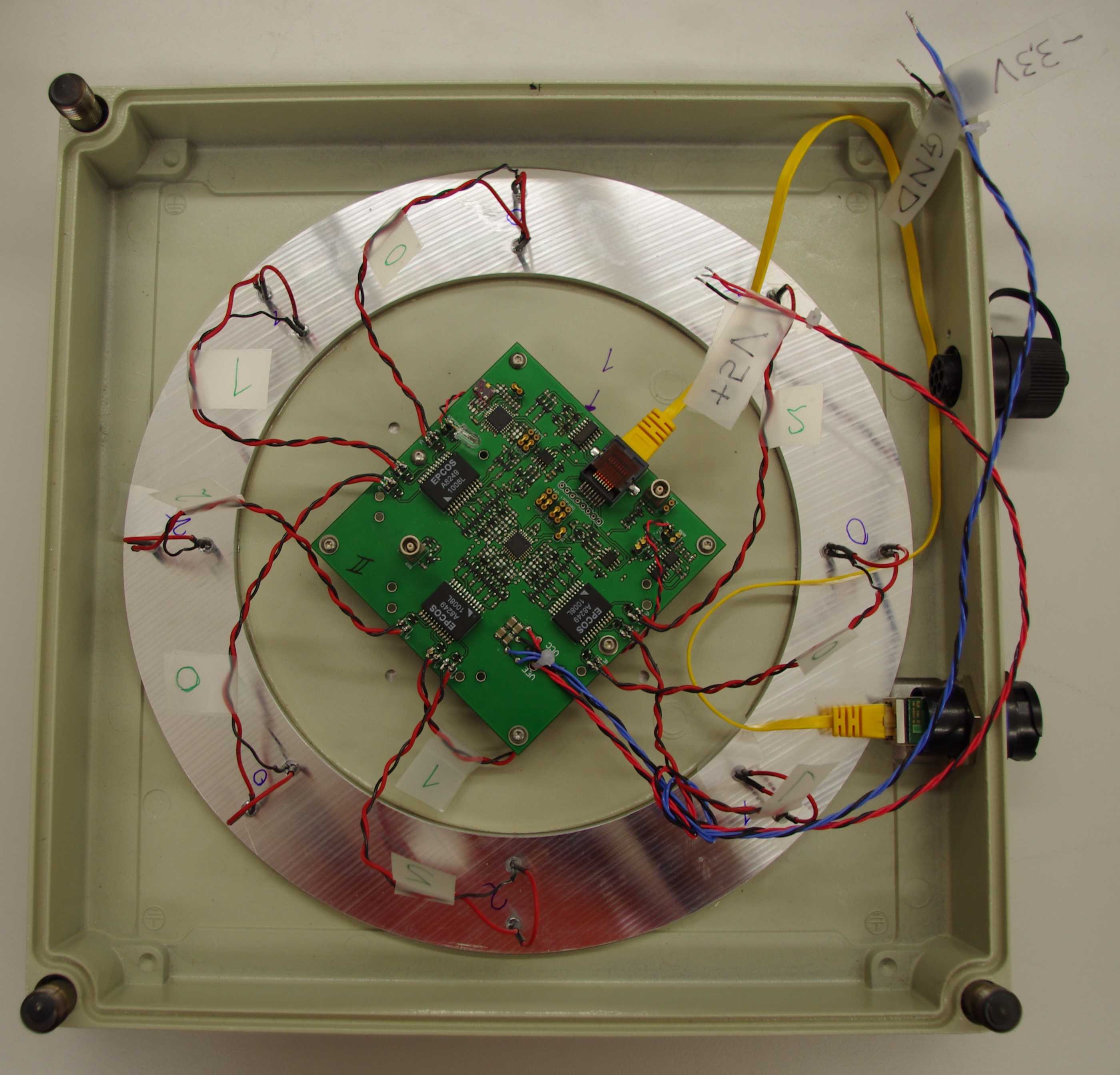}
    \caption{Photographs of the external light pulser. Left: Front view
      showing the geometrical configuration of the LEDs during a test
      installation in the laboratory; at the bottom also
      the connectors for power and control can be seen. Right: The electronics
      board inside the box supplying and stabilizing the LEDs.}
    \label{fig:external_light_pulser}
\end{figure}

The external light-pulser features a total of 18 Light Emitting Diodes
(LED) from Avago Technologies, type HLMP-CB-1A-XY, with a dominant wavelength
of 470\,nm, i.e.\ blue light, and a viewing angle of 15\textdegree{}
defined as the full width half maximum of their radiation
pattern. Both, the electronics as well as the LEDs are mounted in a
commercial aluminum die-cast case. In
figure~\ref{fig:external_light_pulser} a front view (left) and an inside
view (right) are presented. The external light-pulser is controlled by
the trigger master board inside the camera (see
section~\ref{sec:camera:electronics}) via four Low Voltage Differential
Signaling lines (LVDS). Standard RJ-45 Ethernet cables are used to
transmit the LVDS signals. During standard data taking, only the two
central LEDs visible in figure~\ref{fig:external_light_pulser} (left)
are used.

The light-pulser electronics features a feedback circuit in order to
stabilize the light yield of the LEDs over the maximum temperature
range expected at the ORM (-10\,\textdegree{}C up to
+40\,\textdegree{}C, approximately). For this purpose, an additional
LED is mounted inside the case coupled to a silicon pin-photodiode
(Hamamatsu S\,5821). Depending on the output of the photodiode, the
current supply of all (18+1) LEDs is regulated. The amplitude set
point is generated by the trigger master board as a pulse-duty factor
signal and transmitted via one of the LVDS lines. An additional
on-board signal generator allows to operate the light pulser without
the trigger master board for debugging purposes.

\newpage
\section{Camera}\label{sec:camera}

\subsection{Mechanical Layout}\label{sec:camera:layout}

\begin{figure}[hb]
    \includegraphics[width=.49\textwidth]{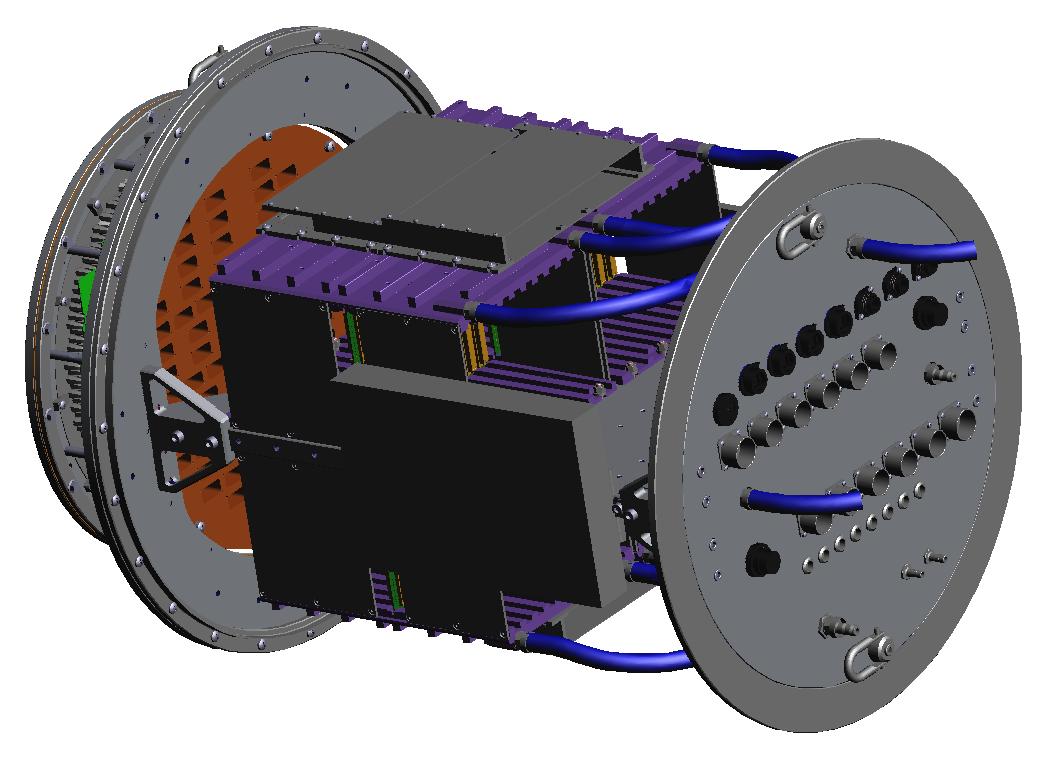}
    \hfill
    \includegraphics[width=.49\textwidth]{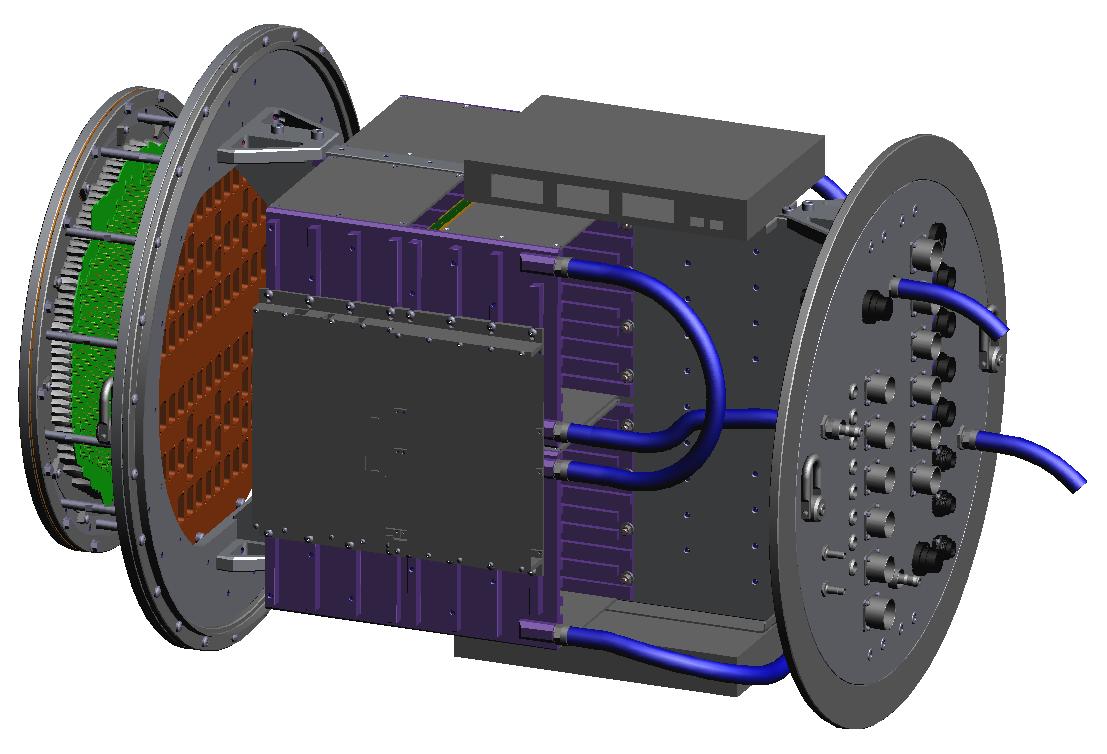}
    \caption{CAD drawing of the camera mechanics without cover
      and shutter. Left/Right: Cooling plate in horizontal/vertical
      position. The sensor compartment can be seen to the left of the
      large electronics compartment in the middle, and the backplane
      to the right. In blue the tubes for the cooling system are
      indicated.}
    \label{fig:CAD_camera}
\end{figure}

The camera has a diameter of 53\,cm and a length of
81\,cm. It consists of two mechanically separated compartments, one
for the photo sensors and one for the readout electronics (see
figure~\ref{fig:CAD_camera}). The sensor compartment has a plexiglass
protection window and light concentrators on the entrance side, with
the G-APDs being glued onto the backside of the concentrators (for
details, see section~\ref{sec:camera:sensors}). To avoid power
dissipation from the electronics to the sensor compartment, an
insulation is installed. This baffle plate is made of 23\,mm Styrofoam
(CORAPAN\textsuperscript{\textregistered} AL\,85),
laminated on both sides by 1\,mm aluminum and equipped with cable
connector boards. The sensor compartment has its own water-tight cover
(2\,mm-thick aluminum cylinder, not shown in
figure~\ref{fig:CAD_camera}) and can be detached from the rest of the
camera.

\begin{figure}[ht]
    \centering
    \includegraphics[height=4.8cm]{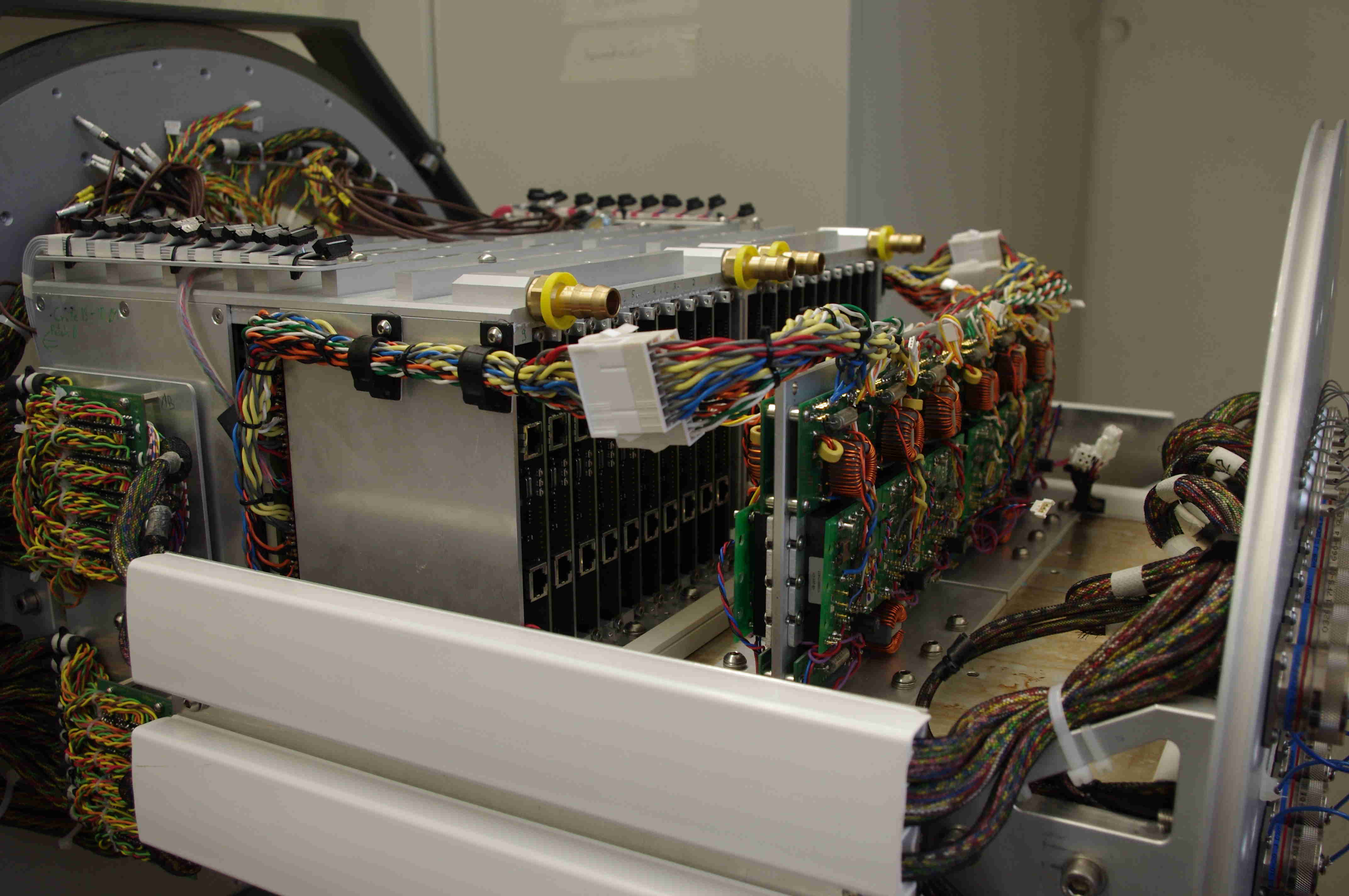}
    \hfill
    \includegraphics[height=4.8cm]{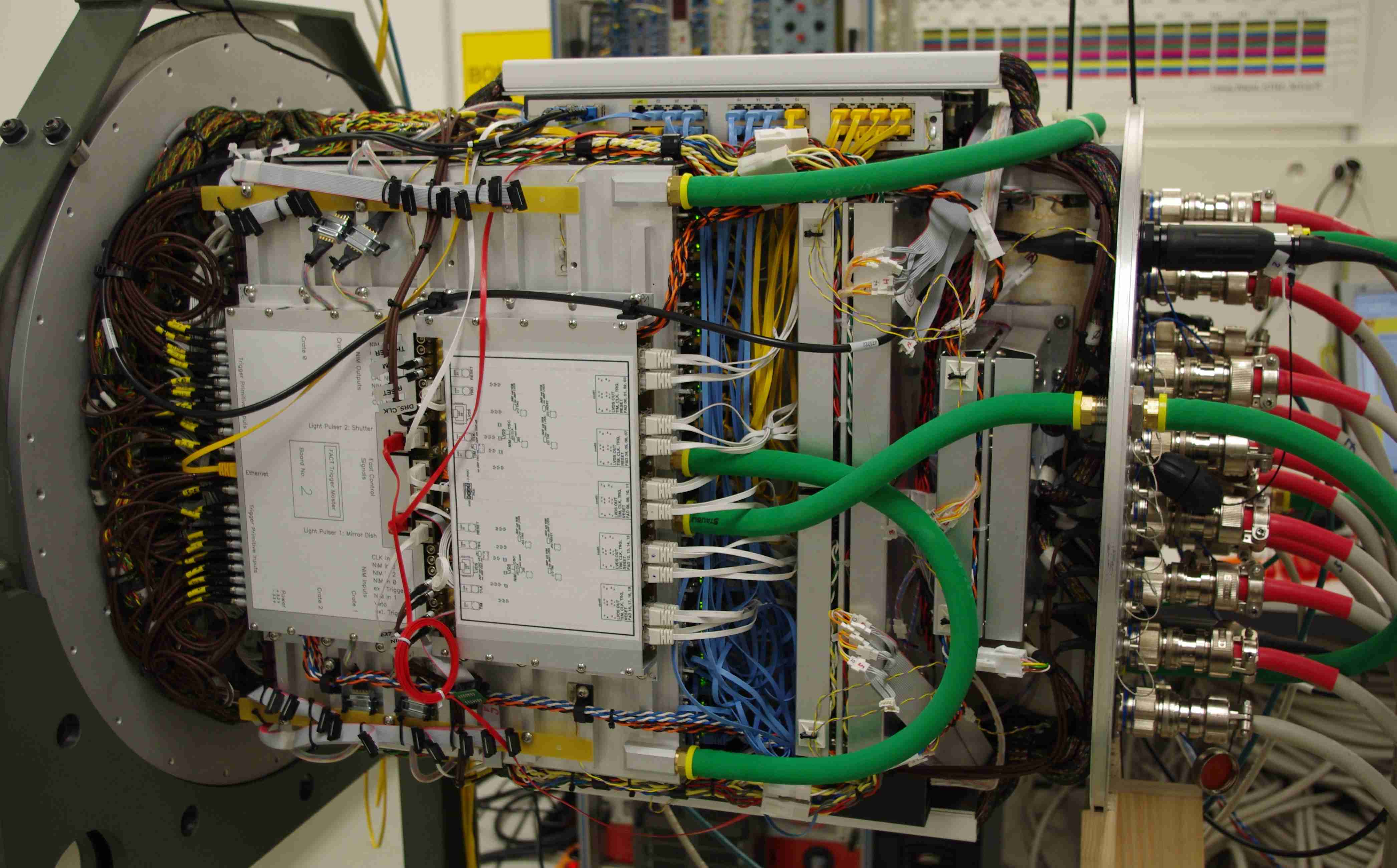}
    \caption{Photographs taken during the construction of the
      camera. Left: Two crates mounted on the (horizontal) cooling
      plate with plugged-in electronics boards and power supplies
      (without cover). Right: Fully equipped and cabled electronics
      compartment (vertical mounting) including cooling tubes; on the
      top one of the network switches is visible, in the foreground
      the cover boxes of the trigger master board and one of the fast signal
      distribution boards; the cables attached from outside to the
      backplane of the camera can be seen on the right.}
    \label{fig:foto_camera}
\end{figure}

Most of the electronics boards are arranged in four water-cooled
crates with 2$\times$10 slots each. These crates constitute the
biggest fraction of the electronics compartment which is subdivided in
two parts by a 24\,mm aluminum plate. This plate serves as mechanical
holding structure as well as heat sink (see also
section~\ref{sec:camera:aux:cooling}). On both sides of the camera,
an Ethernet switch is installed. In the back part of the camera, the
internal power supplies, two on each side, are installed. Four more
electronics boards with individual aluminum covers are mounted on top
of the crates. All aluminum pieces of the camera are alodined to
protect them against corrosion and to keep their thermal and electrical
conductivity. Figure~\ref{fig:foto_camera} (left) shows the camera
during the construction phase with the cooling plate in horizontal
position. For stability reasons, the camera is mounted such that for
the parked telescope the cooling plate is oriented in vertical
position (see figure~\ref{fig:foto_camera}, right). Similar to the
sensor compartment, also the electronics compartment has a 2\,mm-thick
cylindrical cover imposed, varnished with white baking paint. All
connectors for power and communication are located on the backside of
the camera. Dust and water tightness according to IP\,67 are ensured
by using o-ring sealings and dedicated cable feedthroughs and
connectors. A Gore-Tex\textsuperscript{\textregistered}
valve is employed for pressure compensation.

In front of the sensor compartment there is a shutter which can be
opened and closed remotely (not visible in
figure~\ref{fig:CAD_camera} or \ref{fig:foto_camera}, see
section~\ref{sec:camera:aux:shutter}). It consists of two halfs, both
equipped with D-shaped PMMA (Polymethyl Methacrylate, plexiglass)
sheets on their inner side. A total of eight LED pairs is glued onto
the edges of each sheet. This is the so-called internal
light-pulser. In contrast to the external one
(c.f.\ section~\ref{sec:telescope:LP}), the corresponding electronics
board is placed inside the camera. The PMMA sheets are sandblasted on
the side facing the camera such that the light of the LEDs is
scattered onto the pixels. This system can be used for test measurements
during daytime when the shutter is closed.

Including the shutter and the internal light pulser, the camera has a
total weight of 152\,kg.

\subsection{Sensor Compartment}\label{sec:camera:sensors}

\begin{figure}[ht]
    \centering
    \includegraphics[height=5.0cm]{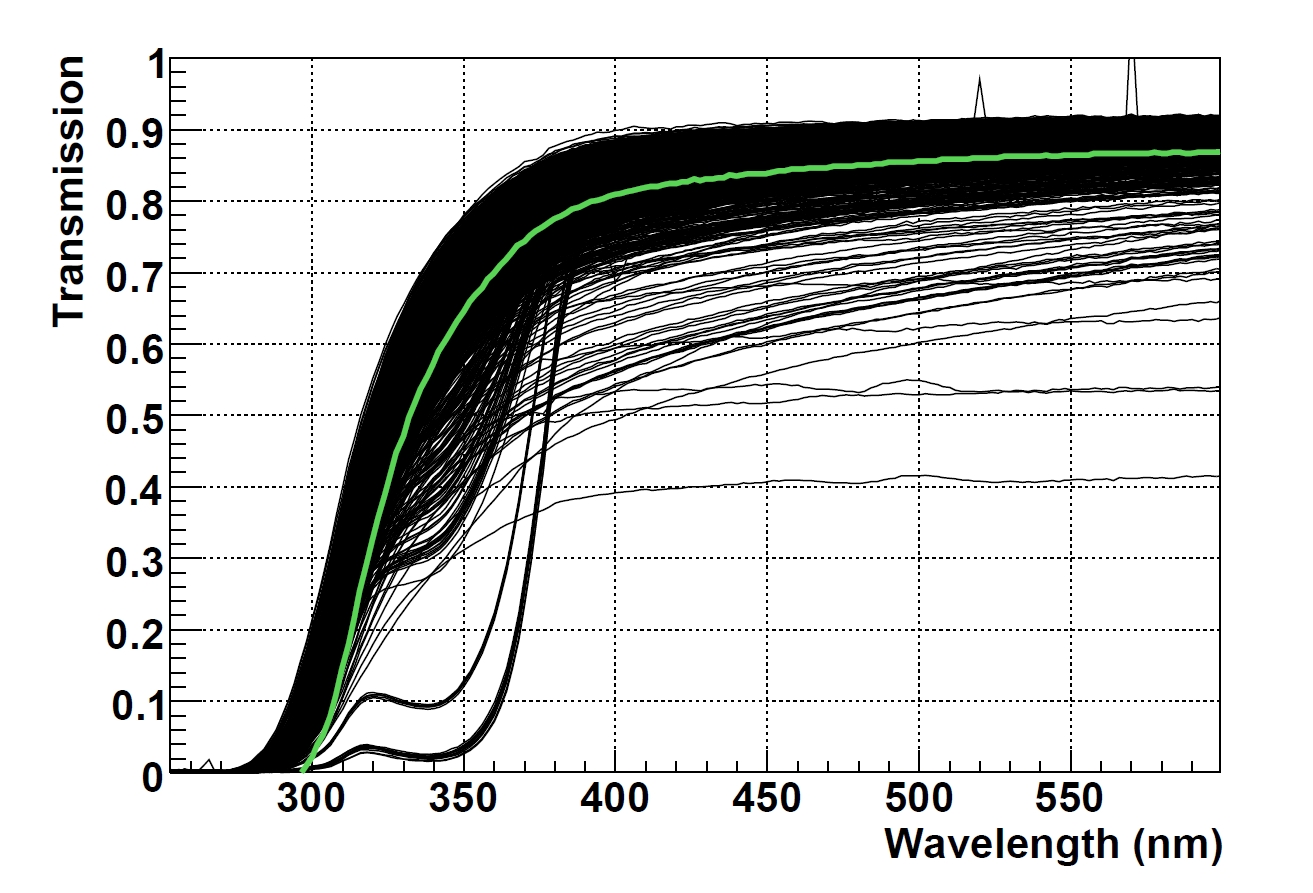}
    \hfill
    \includegraphics[height=4.8cm]{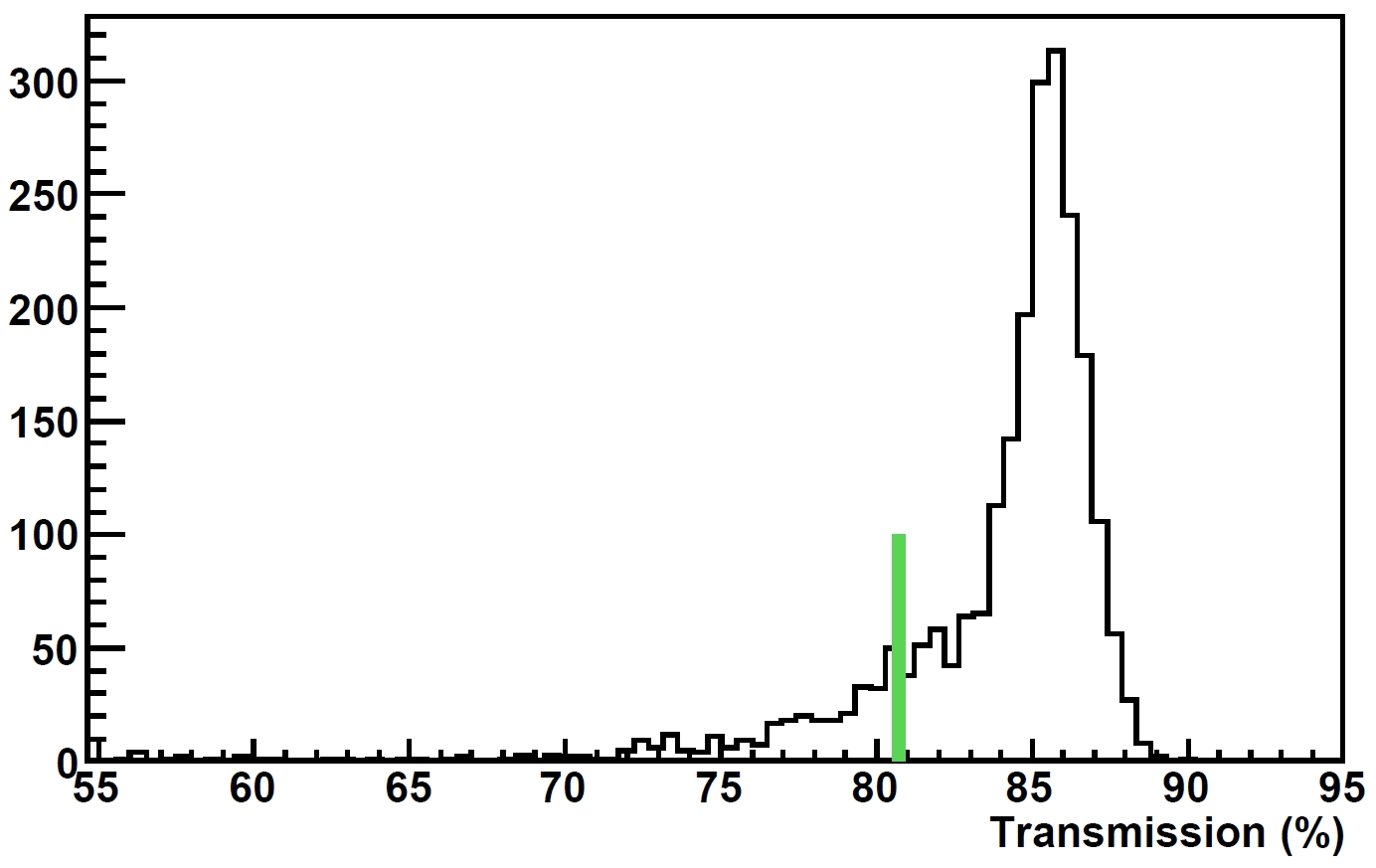}
    \caption{Spectral transmission of the 2344 PMMA light concentrators
      considered for the construction of the camera. The measurements were
      performed with a PerkinElmer LAMBDA 900 spectrophotometer for
      wavelengths between 250\,nm and 600\,nm in steps of 2\,nm. Left: Each
      black line shows the result for one concentrator ($\pm$1\%
      uncertainty). The small jumps are due to measurements errors (short
      exposure to ambient light during one step) and can be ignored. The green
      curve represents the cut-off which was used to reject concentrators of
      poor quality. Right: Projection of the measurement results for a
      wavelength of 400\,nm (zoomed, cut-off line in green). The measurements
      include Fresnel reflection at both, entrance and exit surface.}
    \label{fig:cone_transmission}
\end{figure}

A total of 1595 G-APDs (Hamamatsu MPPC S10362-33-50C \cite{mppc08})
were purchased for the project (not only for the camera, but also for test setups).
The sensors have an active area
of 3\,mm$\times$3\,mm evenly subdivided into 3600 cells. The active
area is enclosed by a ceramic packaging and protected by a thin
(typically 0.3\,mm) epoxy-resin layer with a refractive index between
1.49 and 1.54.
With the resin, the sensitivity for incoming photons has a lower
cut-off around 320\,nm, matching approximately the cut-off introduced
also by the light concentrators (see figure~\ref{fig:cone_transmission}). The dark count rate per
sensor is of the order of a few MHz at room temperature, thus below
the NSB rate even during darkest nights
(c.f.\ section~\ref{sec:intro}). The G-APDs are operated at a gain of 7.5$\cdot
10^5$, with a peak PDE of $\sim$33\% between
450\,nm and 500\,nm and a crosstalk probability of $\sim$13\%. The latter is
defined as the probability that a primary avalanche will
trigger a second or more cells (see e.g.\ \cite{otte09}). Provided that an incoming photon
is detected by the sensor, and assuming that there is no crosstalk or
saturation, one avalanche is equivalent to a single photon signal (photon equivalent, p.e.).

To get a homogeneous response over the camera, the G-APDs were sorted
according to their nominal operation voltage. Those requiring a very
low or very high voltage were not used. After this selection step,
1535 sensors remained with a voltage range from 70.64\,V to 71.63\,V
(for an operation at 25\,\textdegree{}C).

In order to increase the photo-sensitive area per G-APD, a light
concentrator is placed in front of each sensor. Additional advantages
are the filling of the dead space due to the packaging and the
shielding of photons arriving from outside the reflector surface.
While for the prototype module of the camera, a simple (truncated pyramid) open
geometry with
reflective walls was chosen~\cite{anderhub09}, the final camera was
constructed using solid concentrators based on total internal
reflection. Due to refraction at the surface. The achieved compression
ratio can be as much as $n^2$ higher than for hollow concentrators, $n$ being
the refractive index. The applied light concentrators have a height of
20\,mm, a hexagonal entrance with a distance of the two parallel sides
of 9.5\,mm and a square exit of 2.8\,mm\,$\times\,$2.8\,mm. The hexagonal
shape has been chosen to exploit the ideal symmetry for the trigger
system and the event reconstruction. The exit area is smaller than the
G-APDs to have some clearance in the positioning.
The shape of the concentrator surface has been optimized
by 3-dimensional ray-tracing simulations for a cut-off angle for
incident photons of 22.5\textdegree{}~\cite{braun09} having a possible
future extension of the reflector in mind. In the simulations a flat
angular acceptance of the sensors is assumed, which is fulfilled
at least up to 60\textdegree{} for the G-APDs employed \cite{tk09}.
For more details on the performance of the light concentrators, see~\cite{BenCones}.

Apart from a higher compression ratio, an additional advantage of
solid concentrators is that their material and the one of the sealing
window can be selected to match each other as well as the refractive
index of the G-APD resin. In this way light loss due to Fresnel
refractions at several surfaces can be avoided. Furthermore, mass
production techniques can be applied, and also complicated shapes are
available.

The applied light concentrators were produced by injection molding,
using a UV-transparent polymethyl methacrylate (PMMA) granulate
(PLEXIGLAS\textsuperscript{\textregistered} 7N\,OQ, refractive index
of 1.49). Together with the industrial manufacturer of the
concentrators (IMOS Gubela GmbH, Renchen, Germany), the quality was
successively increased until the demanding requirements regarding
optical and surface properties were met. Altogether ten batches were
produced, where the first four were prototypes and the last six were
used for the construction of the camera. Since even for these six
batches the quality was not stable enough, it was necessary to measure
the spectral transmission of 2344 concentrators with a
spectrophotometer and to define a cut-off criterion. As a threshold curve
the lowest measured point at each wavelength for the best 25 samples
were used, further subtracting 3\% absolute transmission. The results
of the transmission measurements are summarized in
figure~\ref{fig:cone_transmission}. Two main features are clearly
visible for the rejected concentrators, namely an absorption shoulder
between 300\,nm and 380\,nm and a generally low transmission for several
concentrators. Both can be explained by problems during the
production process, like contamination with leftovers from other
(UV-absorbing) PMMA composites or changes in the cooling process. It
should be noted that the measurements include the losses due to
Fresnel reflections at the entrance and exit area of the concentrators
(transition from air to PMMA and vice versa), which account for about
two times 4\%. For a camera pixel the Fresnel losses are reduced,
because the refractive index of PMMA is very similar to the one of the
resin on top of the G-APDs (see above).
%

\begin{figure}[b]
    \centering
    \includegraphics[height=5.4cm]{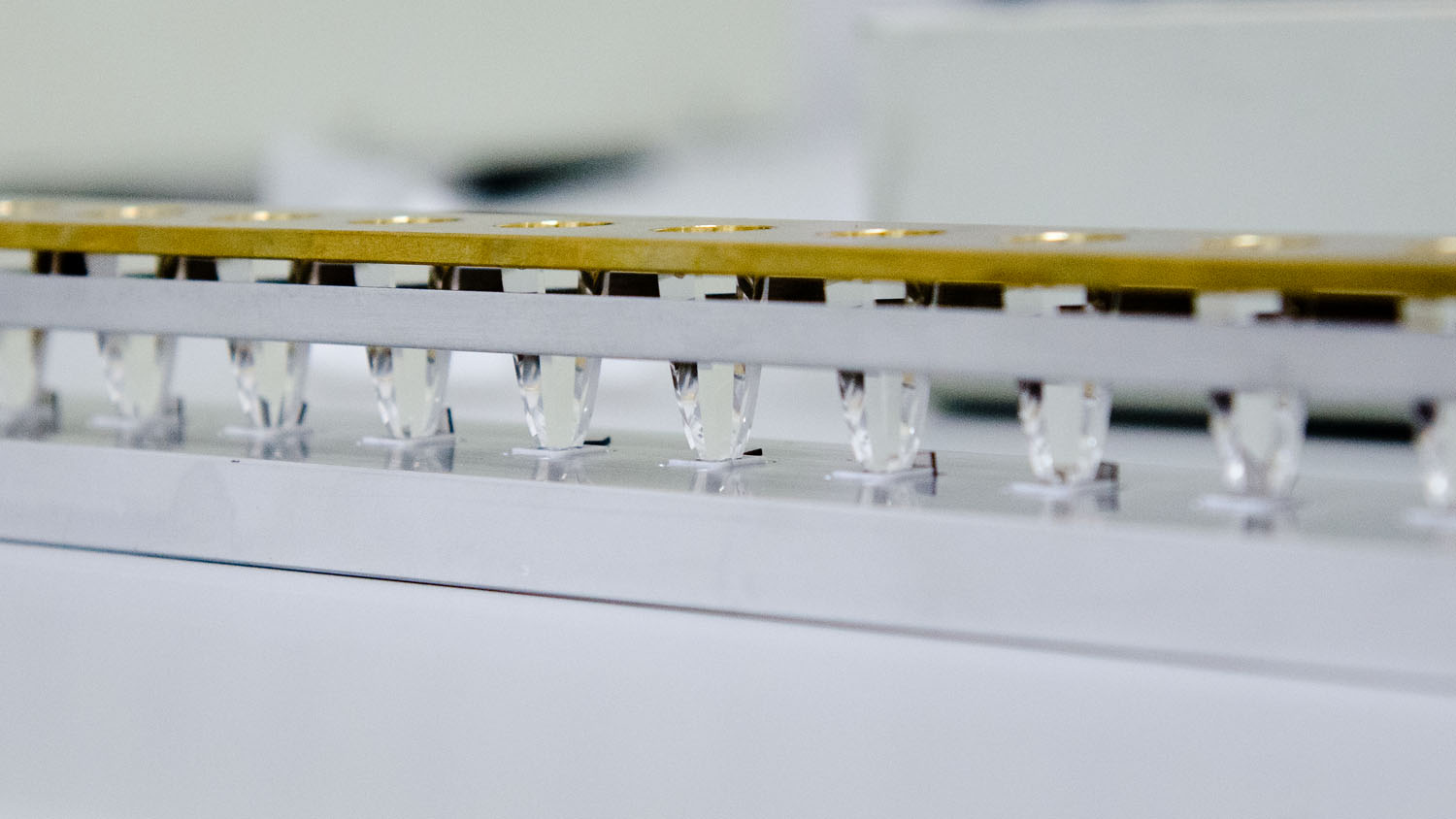}
    \hfill
    \includegraphics[height=5.4cm]{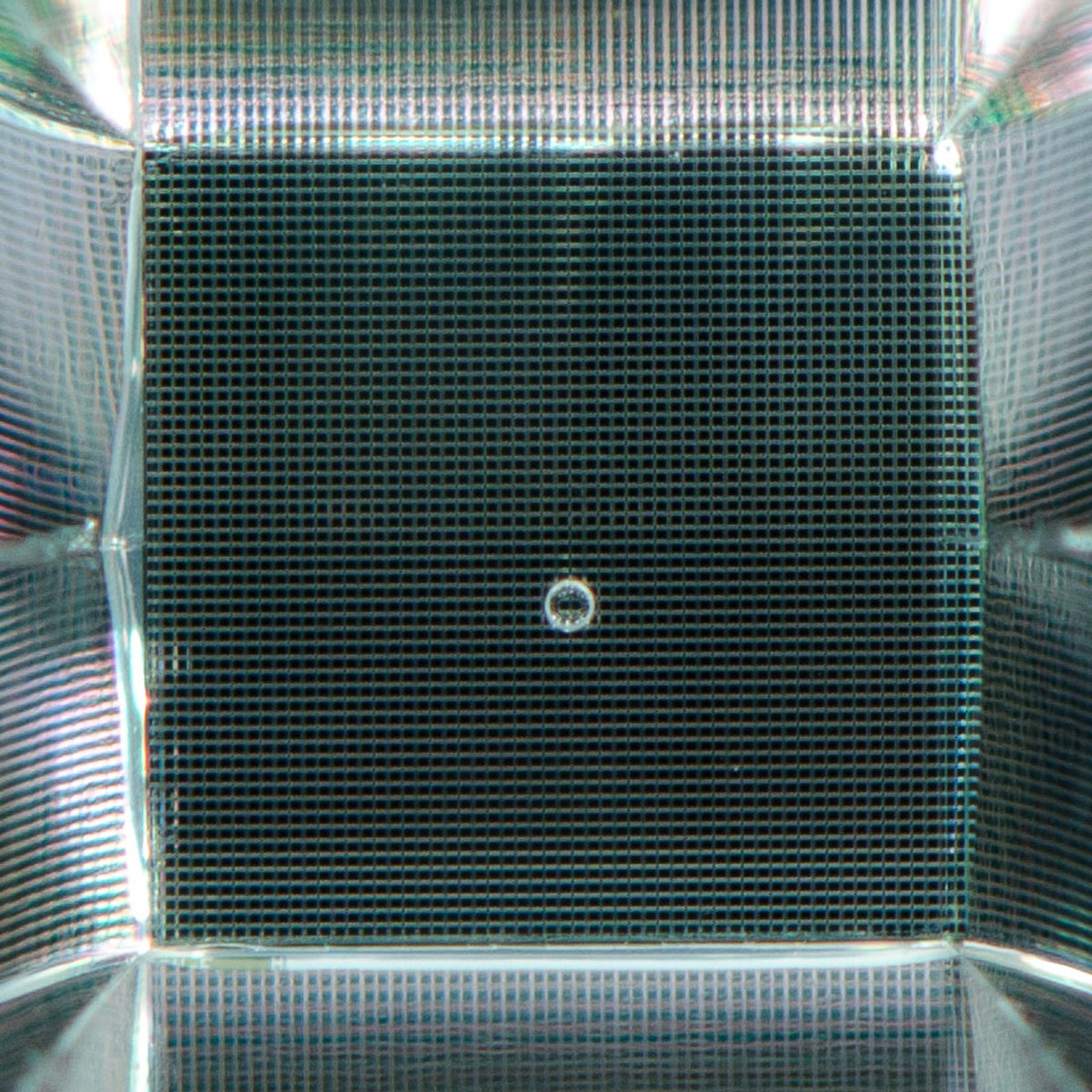}
    \caption{Left: Photograph of the alignment tool to glue the light
      concentrators onto the G-APDs. The sensors are fixed in
      custom-fit sinkholes on the bottom (white ceramics packaging
      visible), while the concentrators are positioned with two
      horizontal metal bars. Right: One of the pictures taken per
      pixel for quality control. The 3600 G-APD cells are visible and
      also part of the concentrator walls. The bright spots at the
      borders are due to reflections. In the middle, an air bubble can
      be seen, affecting 16 cells. Since this is just 0.4\% of the total area,
      this pixel was accepted. Also it should be noted that only a
      fraction of the photons hitting the bubble really get lost.}
    \label{fig:pixel_assembly}
\end{figure}

To each of the 1535 selected G-APDs a light concentrator fulfilling
the transmission requirements was glued. A two-component glue was used
(EPO-TEK\textsuperscript{\textregistered} 301), which has a refractive
index similar to the one of the protection resin ($\sim$1.5) and which
is transparent down to 300\,nm. The glue was first placed on the
concentrators with a high precision dispenser, and the concentrators
then attached to the G-APDs using custom-made alignment tools (see
figure~\ref{fig:pixel_assembly}, left). Afterwards, both, a load test
and a functionality test were performed. During the load test the
pixels were fixed on the G-APD side (in horizontal position) and a
200\,g weight was put on the opposite side on the concentrator.

For the functionality test, the G-APDs were supplied with their nominal
voltage and illuminated with a nano-second light source at a
wavelength of 450\,nm to check for a signal. While two pixels failed
the load test, none of the remaining ones failed the functionality
test. In addition, a high-resolution picture of the sensor area as
seen through the concentrator was taken with a macro-lens (105\,mm
focal length; see figure~\ref{fig:pixel_assembly}, right). In this way
glueings of poor quality, e.g.\ with many or large air bubbles, could
be spotted. Inspecting these pictures, 24 more pixels were rejected.
Thus 1509 pixels were available for the camera construction.

\begin{figure}[hb]
    \centering 
    \includegraphics[height=4.22cm]{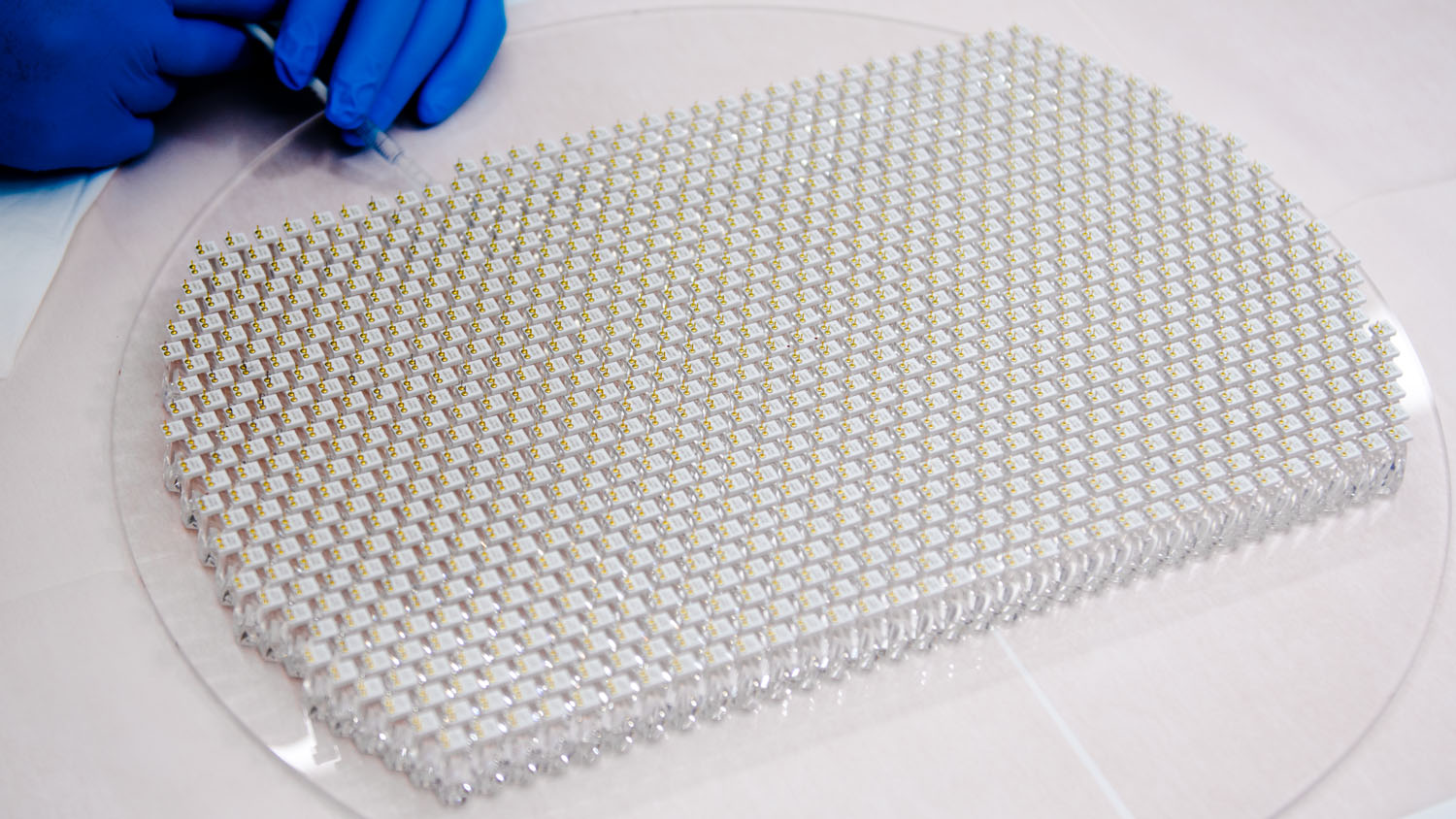}
    \hfill
    \includegraphics[height=4.22cm]{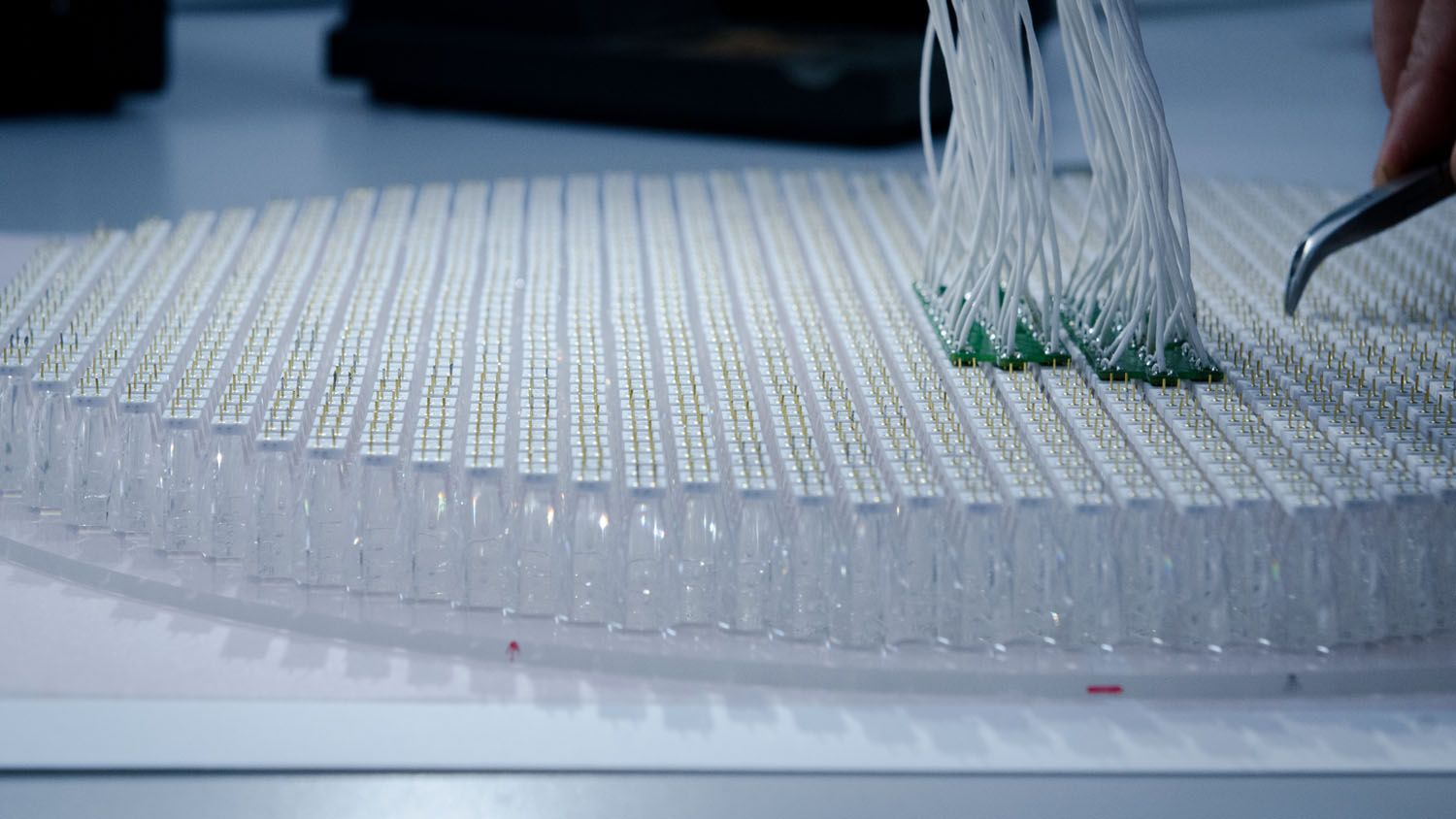}
    \caption{Photographs documenting two work steps during the
      assembly of the sensor compartment. Left: Glueing of the pixels
      onto the camera window. Right: Soldering of the PCBs for the
      signal and bias-voltage connection on the G-APD contacts (facing
      upwards).}
    \label{fig:window_assembly}
\end{figure}

In the next assembly step, 1440 pixels were glued to the camera
window which is optically identical to the concentrators.
For the glueing the concentrators were placed face
down on the window, and the glue
(Acrifix\textsuperscript{\textregistered} 1R 9019 Solar) was injected
at the side (see figure~\ref{fig:window_assembly}, left). After some
unsatisfying first attempts the manufacturer (R\"ohm Evonik GmbH,
Darmstadt, Germany) was consulted and some
Acrifix\textsuperscript{\textregistered} 1R\,9016 Solar was admixed. This
avoided air bubbles because the glue was sucked in more
homogeneously by capillary forces.

\begin{figure}[t]
    \centering
    \includegraphics[width=\textwidth]{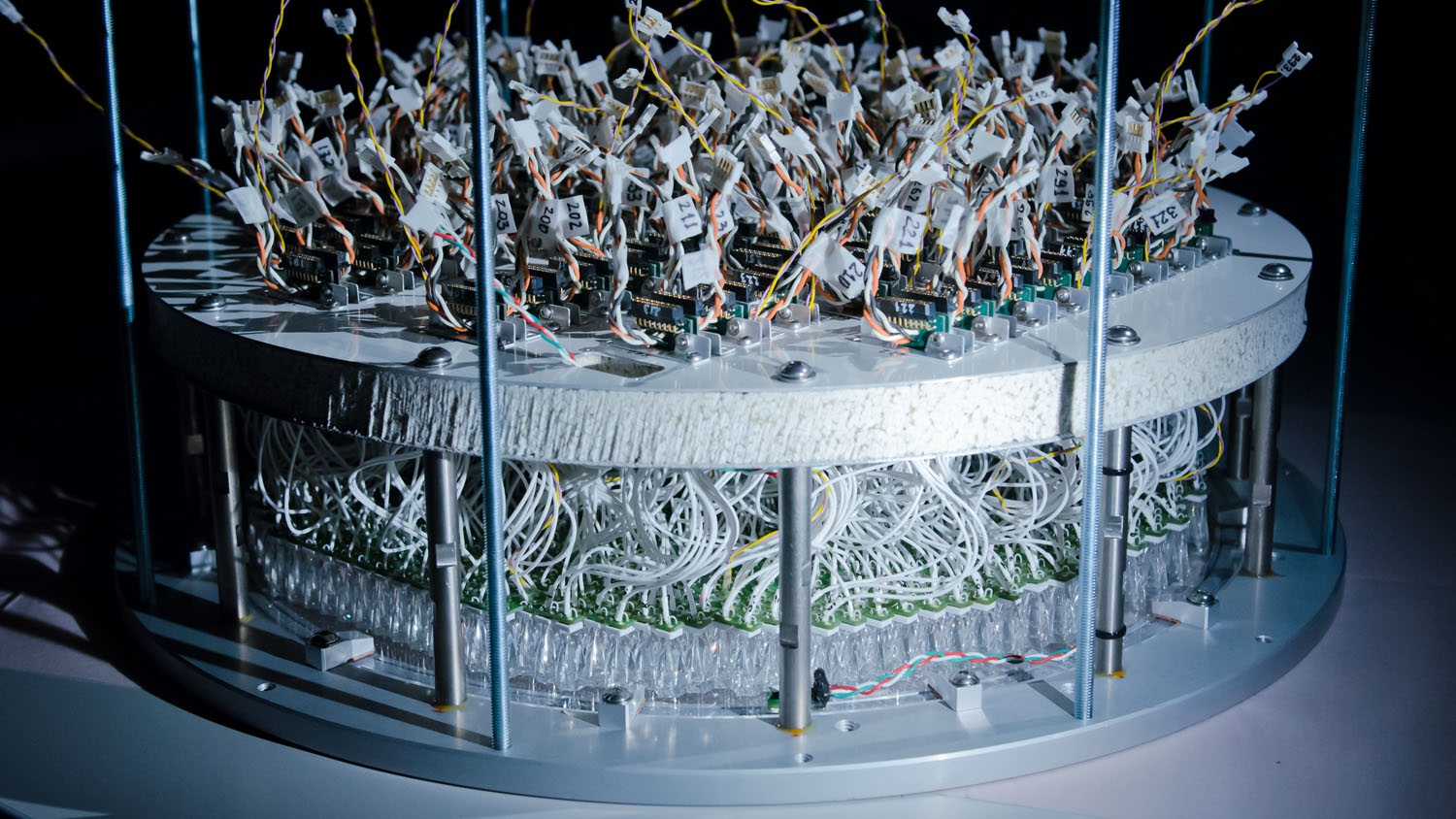}
    \caption{The assembled sensor compartment. From bottom to top the
      front flange with the entrance window, the light concentrators,
      the G-APDs (white packaging), the PCBs (green) for the co-axial
      cables (white), and the insulation plate can be seen. The black
      connectors on top are for the G-APD signals, the three-pole
      cables (white-black-orange) for the bias supply. Other colored
      cables (yellow-purple and red-green-white) connect the
      temperature and humidity sensors. For the final mounting the
      sensor compartment is protected by a cover (replacing the thread
      bars).}
    \label{fig:sensor_compartment}
\end{figure}

Since four or five G-APDs share the same voltage supply channel (c.f.\ 
section~\ref{sec:overview}), the pixels were sorted before being glued to
the window. The resulting spread in breakdown voltage is $\le$~10\,mV for $94\%$ of
the 320 groups and the remaining 6\% were mounted at the border of the camera.
In order to avoid cable tensions on the G-APDs, small Printed
Circuit Boards (PCB) were soldered to their back side (see
figure~\ref{fig:window_assembly}, right). The co-axial cables for the
bias voltage and the signal transmission are connected to these PCBs
(one cable per sensor, nine sensors per
PCB). Figure~\ref{fig:sensor_compartment} shows a photo of the fully
assembled sensor compartment (without cover). The PMMA window is fixed
to an aluminum flange to which also several distance pieces are
attached.  These bars support the insulation plate introduced in
section~\ref{sec:camera:layout}, which includes boards for
electrical connections. While the co-axial cables coming from the
G-APDs are plugged at one side of the connector boards, on the other
side (in the electronics compartment) plug-connections for the signals
and the bias voltage are provided. Each connector board serves a group
of nine pixels, thus two voltage and nine signal channels. In this way
the sensor compartment can be de-attached from the rest of the
camera. To monitor the conditions near the G-APDs,
31 temperature and two humidity sensors have been installed (IST
P1K0.520.4W.x.010 and Honeywell HIH-4031-0015, respectively).

\subsection{Readout and Trigger Electronics}\label{sec:camera:electronics}

\begin{figure}[ht]
    \centering
    \includegraphics[width=\textwidth]{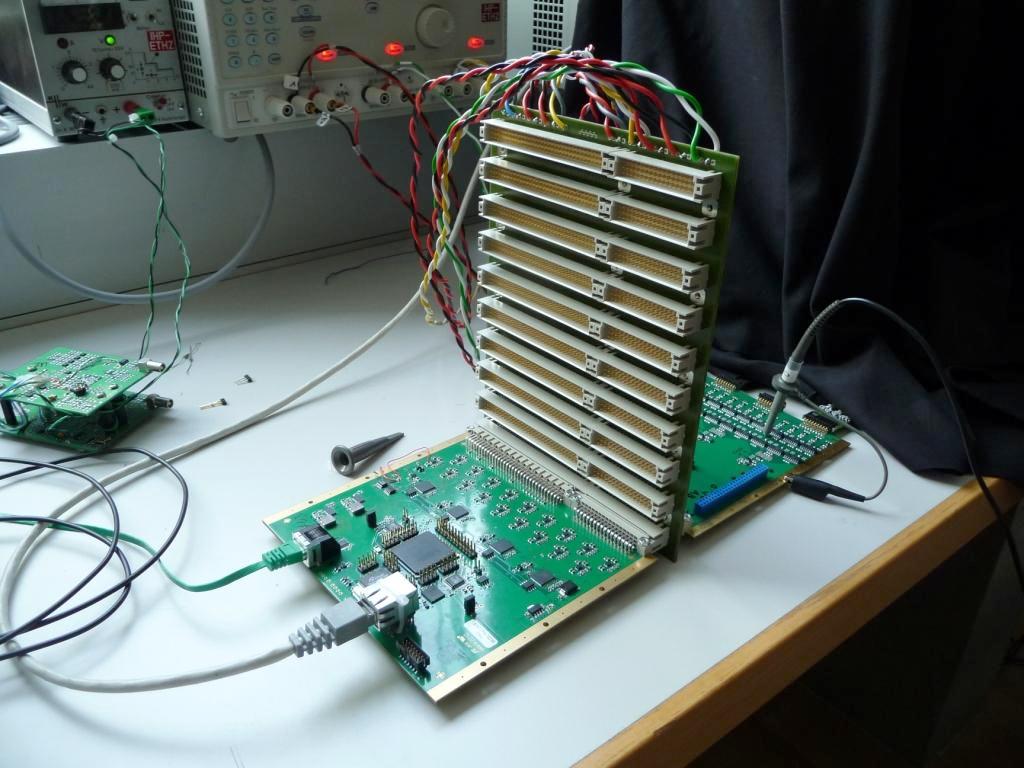}
    \caption{Laboratory setup with the midplane of the electronics
      crates and one preamplifier board (right) and one digitizer
      board (left) plugged in the first slot. On the top, temporary
      cables for low-voltage supply and slow-control communication are
      connected.}
    \label{fig:midplane}
\end{figure}

The analog signals from the sensor compartment are processed and
digitized inside the camera by means of custom-made electronics
boards. One pixel corresponds to one readout channel, thus in total
there are 1440 channels. Most of the electronics is realized on 40
board pairs, which are mechanically integrated in four crates as
introduced in section~\ref{sec:camera:layout}. Each pair consists of a
preamplifier and a digitizer board and serves 36 channels. The two
PCBs of a pair are connected via one midplane per crate, which carries
in addition supply voltages and RS\,485 communication
lines. Figure~\ref{fig:midplane} shows a photograph of the midplane. In
order to initiate the digitization and the transfer of the event data, the
digitizer boards receive a logical signal provided and distributed by
the trigger system. For an overview, see also \cite{twepp12}.

\subsubsection{Preamplifier and Trigger Signal Conditioning}

The preamplifier board features 36 channels grouped in four patches
reflecting the structure of the trigger system and the digitizer boards.
Its first stage is a grounded-base circuit (npn transistor, BFR\,182)
with an input impedance of 25\,$\Omega$. At the collector resistance (200\,$\Omega$) the
sensor current is converted into a voltage, which is further amplified (OPA\,3691, Texas Instruments)
giving after line matching a conversion gain of 0.45\,mV/$\mu$A
at about 200\,MHz bandwidth. This corresponds to a single-avalanche amplitude of 2.25\,mV at the
50\,$\Omega$ input of the digitizer board. For trigger conditioning, the preamplifier signal is used as well.
First a linear addition of the signals from one patch is performed (OPA\,691, Texas Instruments), with the possibility to exclude
individual pixels (e.g.\ due to noise). Cable-based
clipping is used to shorten the length of the 9-fold sum signals to 10\,ns before they
are fed into a comparator (LMH\,7220, National Semiconductor). At the input of the comparator an avalanche
is represented by a voltage of 9\,mV. The
comparator signal from each of the four patches is sent to a trigger
unit realized as a mezzanine card on the preamplifier board. This unit provides also  individual pixel disable
signals and the comparator thresholds.

\subsubsection{Trigger Unit}

The trigger unit is equipped with an on-board field-programmable gate
array (FPGA, XILINX Spartan-3AN family, XC3S400AN-4FGG400C). It
controls an 8-channel digital-to-analog converter (DAC, LTC\,2620)
with a resolution of 12\,bit and a full range of 2.5\,V (270
photon equivalents) to set the comparator thresholds (15 DAC
counts/p.e.) provided to the preamplifier board. The four comparator
output signals from the latter are summed at a time, and the summed
signal serves as input for an N-out-of-4 logic. The threshold level
$N$ is defined by a discriminator, whose threshold is also provided by
one of the FPGA controlled DACs. During standard physics operation,
the N-out-of-4 logic is operated as 1-out-of-4. Since this
discriminator logic requires a minimum length of the signal of the
order of several nanoseconds, by design, the second trigger stage acts
as a filter on electronics noise. The digital output signals of
all 40 trigger units are sent via coaxial cable to the trigger master
board making the final decision and distributing the trigger signal
to the digitizer boards. To monitor the rate of
all trigger channels, a counter for all comparator and discriminator
outputs is implemented. This is essential in order to adapt the threshold
for single more noisy channels due to e.g.\ bright
stars in the field-of-view during operation. The counter readings as
well as the command values for the setup of the trigger unit are
transmitted to/from the trigger master board via an RS\,485 serial
connection.

\subsubsection{Trigger Master Board}

The main purpose of the trigger master board is to generate the final trigger
signal based on the input from the 40 trigger units, and to provide
a common reference clock for all sampling
chips on the digitizer boards. Its core component is an FPGA (Xilinx Spartan-3 family,
XC3SD3400A-4FGG676C) controlling all sub-components and providing a
100\,Mbit Ethernet connection to the counting house through a Wiznet W\,5300 chip.
A trigger decision is taken when $N$ out of the 40 primitives have a
rising edge within a time-window adjustable from 8\,ns to 68\,ns in
steps of 4\,ns. During physics data-taking this majority coincidence
is operated with $N=1$ and the window is set to 12\,ns. It is avoided
to process the trigger signals in the FPGA on the trigger units to
keep their individual jitters as small as possible. The jitter on the
trigger output from the trigger master board FPGA is a few nanoseconds and can
be neglected for the global readout time.

In addition to the physics triggers, the trigger master board can generate
pseudo-random triggers, as well as triggers for the internal and
external light-pulser with a rate between 1\,Hz and 1\,kHz. After
sending a trigger to one of the light-pulsers, the multiplicity $N$ is
changed for a short time-interval to e.g.\ 25 to avoid interference
with physics triggers. Furthermore, two external trigger inputs and a
veto input are available. All trigger outputs can be switched on and
off individually such that, for example, random and light-pulser
events can be interleaved during data-taking.

After a customizable delay, a rectangular signal, called time-marker,
is created and distributed to all digitizer boards where it is mixed
into one channel of each sampling chip. Having the delay adjusted such
that this marker is sampled at the end of the readout window, it
serves as cross-check for individual delays between boards and chips.
After a second adaptable delay, the trigger decision is propagated as
logic signal to the data-acquisition boards to stop the analog
ring-buffers continuously sampling the data for readout. Both delays
are implemented in the FPGA. During digitization, and until sampling
is restarted, the data acquisition boards emit a busy signal
preventing further triggers during that time.

Just after a trigger decision, a bit pattern containing a trigger
counter and the source which generated the trigger is sent via a
serial RS\,485 bus, one per crate, to the data acquisition boards.
There it is included in the raw data stream. A custom dead-time is
set, to ensure new triggers are rejected until the transfer is
complete.
The trigger counter is stored and can be readout via Ethernet together
with counters for the total run-time and on-time. The latter takes into account the
busy-time of the DAQ boards, the dead-time and also the time to wait for light-pulser triggers.
It thus measures the overall time when the camera is ready to accept triggers.

Additionally to the trigger signal, the trigger master board provides the
reference clock for the digitizer boards as 1/\nth{2048} of the sampling
rate. In order to generate this signal with a jitter below 100\,ps,
a clock conditioner (National Semiconductor LMK\,03000) is used.
This device also provides the possibility to change the sampling rate
within certain limits. A second
output of the clock conditioner is used to generate a square signal,
which can be distributed as substitution of the time-marker to
calibrate the timing behavior of the analog pipeline chips on the digitizer boards. In order to
distribute the fast signals (trigger, reference clock and
time marker) the LVDS standard
and slim Wirewin Cat.6 cables with RJ-45 connectors are used. Two dedicated distribution
boards were designed for this purpose. They comprise two ten-fold
fan-outs (ON Semiconductor MC100LVEP111) for every signal, and
introduce only a low jitter ($<20$\,ps) and skew ($<250$\,ps).
Auxiliary input and output signals are distributed as NIM (Nuclear
Instrumentation Module) logic levels. The configuration of all 40
trigger units is done through the trigger master board and communicated via
a dedicated RS\,485 bus.

\subsubsection{Digitization}\label{sec:camera:electronics:FAD}

The digitization of the 1440 G-APD signals is done on
40 boards, each comprising 4$\times$9 channels.
These digitizer boards have five important parts: input buffers, analog pipeline chips,
analog-to-digital converters (ADC), an FPGA, and an Ethernet chip.
At the input buffers the single-ended preamplifier signals are converted to differential
ones using fully differential amplifiers (ADA\,4950, Analog Devices) set to a gain of two at a bandwidth of 300\,MHz.
This chip allows also to feed in signals for timing calibrations,
DC levels for amplitude calibrations, and an adjustable common-mode voltage.
The latter shifts the signals into the optimal range for the pipeline chips having
a dynamic range of about 1\,V.

Sampling of signals is based on the Domino Ring Sampler (DRS\,4)
chip, developed at the Paul Scherrer Institute \cite{ritt10}. Each chip comprises 9$\times$1024 capacitors and samples continuously the nine input voltages. The values are stored in pipelines until a trigger stops the sampling.
Any number of samples can then be digitized. The system can be operated
with a sampling frequency of 0.7\,GHz -- 6\,GHz. During standard data taking 2\,GHz are used and
300 values (150\,ns) are read out. In this way also large
signals from the light pulsers are completely contained in the readout window. During calibration
measurements, necessary to calibrate the properties of the capacitors
of the DRS\,4, all 1024 samples are digitized. In order to calibrate the
gain of each capacitor, a voltage level is fed into the
DRS\,4 input via the differential amplifier. This voltage is produced by a
16\,bit digital-to-analog
converter (DAC, LTC\,2600, Linear Technology). For
more details on the DRS calibration, see \cite{ritt10} and references
therein.

All 160 DRS\,4 chips employed in the electronics are locked using
their internal phase-locked loop to a common frequency generated by
the trigger master board. In this way, a precise relative timing between all
channels of about 300\,ps is possible \cite {ritt10} (depending on the sampling
frequency), provided that the fixed-aperture jitter of the DRS\,4 is
calibrated. This timing calibration is performed by using a 250\,MHz
clock signal, generated by the trigger master board. For dedicated runs, this
signal can be capacitively coupled into channel nine of each DRS\,4
chip. Alternatively, a high-precision time marker, also generated by
the master board, can be imprinted on the
analog signal of every ninth channel. In this context, it serves as
cross-check for the timing calibration.

Following a trigger, 12 bit analog-to-digital conversion (AD\,9238, Analog Devices) at 20\,MHz is started.
The differential outputs of the DRS\,4 have a gain of two, thus allowing in principle for a dynamic range of 2\,V at the ADCs.
However, the dynamic range is reduced to 1.8\,V by an intended baseline shift and a compensation for individual DRS\,4 offsets. All digital settings are controlled by an FPGA (Xilinx Spartan-3 family, XC3SD3400A-4FGG676C), i.e.\ the sampling mode, the serial conversion mode, the calibration options, and the DAC values.
Another important task of the FPGA is to organize the communication and data sending. Each digitizer board has an Ethernet interface (Wiznet W\,5300 chip) implemented for
this purpose.

\begin{figure}[t]
    \centering
    \includegraphics[trim=0.7cm 0cm 2.3cm 1.3cm,clip=true,width=\textwidth]{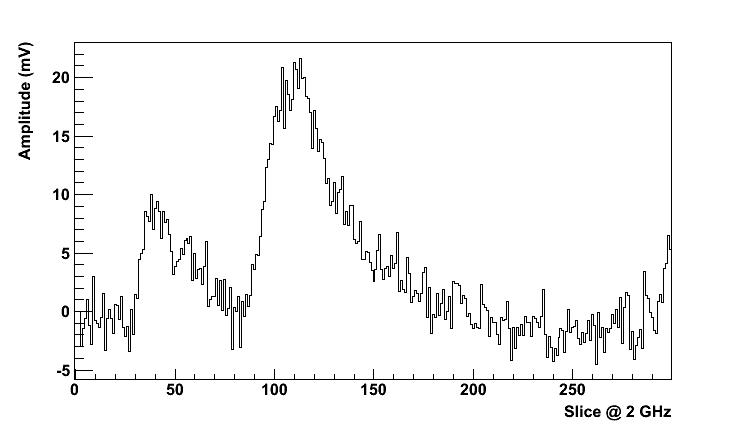}
    \caption{Example of an event as recorded by one readout
      channel. The data were taken with closed shutter and using a
      random trigger. A single and a double-avalanche signal are visible.
      On the x-axis the sample position within
      the DRS\,4 pipeline (sample number after trigger) is given in steps
      of 0.5\,ns. On the y-axis the amplitude is shown (1800\,mV full scale).}
    \label{fig:dark_counts}
\end{figure}

Figure~\ref{fig:dark_counts} illustrates the performance achieved with
the readout electronics. Shown is one frame, i.e.\ 150\,ns, of a
single channel recorded with the corresponding G-APD on nominal
voltage and the shutter of the camera closed. A random trigger has
been used for data taking. The DRS calibrations have been calculated 
using calibration runs taken shortly before the data run and have been applied
offline to the data.
Two signals due to dark counts are visible, where the
double-avalanche is probably due to optical cross-talk within the photo
sensor. While the noise level is about 2\,mV root-mean-square, the
peak amplitude per avalanche is about 10\,mV.
The total gain of the signal chain is
1.8\,mV/$\mu$A.
%
\subsection{Power, Cooling and Auxiliary Systems}\label{sec:camera:aux}


\subsubsection{Power Conversion and Distribution}\label{sec:camera:aux:power}

Three-phase AC power is received from the MAGIC \cite{magic} power
lines. A 3-phase filter followed by a main choke provides power to the
drive cabinet. In parallel to the drive, the three phases are
filtered individually, separately serving the computer infrastructure,
the camera and the cooling system. The camera uses three Agilent AC-DC
supplies, one for the G-APD bias supply (NS\,5769A) providing 85\,V,
another one for the interlock system and the heaters providing 24\,V,
and the main supply (NS\,8737A) for all camera systems providing 48\,V
at about 12\,A. Two 45\,m long cables bring the power to the patch
panel (backplane) of the camera. Inside the camera DC-DC converters
(VICOR VI-J300 series) are used for power conditioning. The output of each
converter is equipped with an adapted filter, mainly a common mode
choke and a low ESR Tantalum capacitor, which is mandatory in order to
achieve the required noise levels. In addition, step-down converters
are used on the digitizer boards
(c.f.\ section~\ref{sec:camera:electronics:FAD}).

All output voltages and currents are monitored
continuously using shunt resistors. For this purpose the corresponding
analog signals are
routed to a slow control board (see
section~\ref{sec:camera:aux:slow}) where they are digitized and
transmitted to the counting hut. The camera houses four
DC-DC converter boards, each serving one of the DAQ crates and two
additional DC-DC converter boards supplying all remaining power
consumers, namely the trigger master board, the two fast signal distribution
boards, the internal and external light-pulser, the slow control board,
and the two Ethernet switches.

The total power consumption of the camera electronics during operation
is about 570\,W. About 100\,W are dissipated in the supply
lines and another 100\,W due to the limited efficiency of the DC-DC
converters, yielding a total of 370\,W consumed by the electronics. An
additional 100\,W are used by the G-APD bias supply systems. During
powering up the FPGAs on the digitizer boards and on the trigger units,
significant extra power is
needed. The resulting load would exceed the limits of the DC-DC
converters (and eventually of the power supply).
Consequently, the digitizer boards in each crate are
inter-connected such that they boot sequentially, one after the other.

\subsubsection{Bias Voltage Supply}\label{sec:camera:aux:bias}

\begin{figure}[t]
    \centering
    \includegraphics[width=\textwidth]{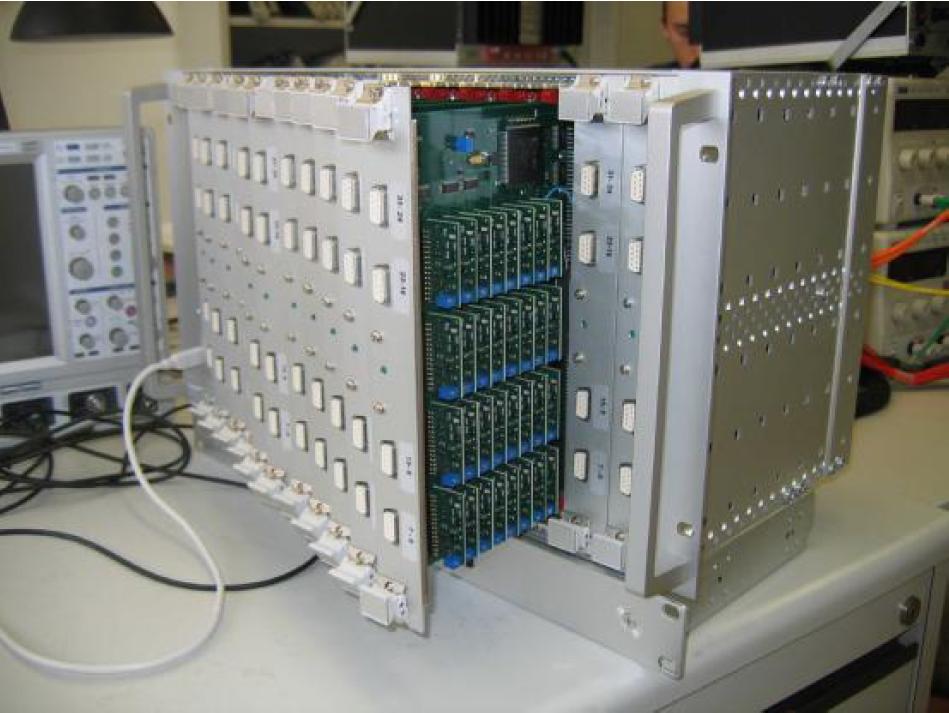}
    \caption{The bias-voltage crate supplying the G-APDs. On the left,
      where the USB cable is plugged, the controller is located. For
      this photograph most of the motherboards have been pulled
      out. On the board in the front the 32 plug-in cards, each
      constituting one channel, are visible.}
    \label{fig:bias_crate}
\end{figure}

The bias supply system provides up to 416 DC voltage
channels, each programmable between 5\,V and 90\,V with 12\,bit
resolution ($\sim$20\,mV step width). Currents of
up to 4\,mA per channel can be supplied and measured with 12\,bit
resolution. In practice 1\% precision is achievable due to
regulation noise. Each channel has an over-current protection circuit,
disabling that channel at the limit of 4\,mA until a reset command is
received. The complete system is built as a double-height VME-like
crate with slots for 14 subunits. While one slot is occupied by a
controller, up to 13 motherboards with 32 voltage channels each can be
plugged. All units are forced-air cooled. For the operation of the sensors
320 voltages channels (ten motherboards) are needed. In
figure~\ref{fig:bias_crate} a photograph of the crate is presented.

A single external voltage supply providing up to 100\,V is employed
(Agilent N\,5769A) from which the output of each channel is
derived. Usually a voltage limit of 85\,V is set for this unit. Each
channel is realized as a plug-in card to a motherboard. A high voltage, high current
operational amplifier (OPA\,454, Texas Instruments) with internal over-temperature and over-current
protection is used, controlled by a 12\,bit digital-to-analogue converter
(DAC). Precise calibration of the output voltage requires adjusting a
potentiometer while measuring the voltage from each card. The
motherboards employ a Programmable Logic Device (PLD) to communicate
with the controller, the DAC and the ADC chip. Also the controller is
based on a PLD and communicates based on a three-byte protocol via a
USB connection with the control computer. The USB connection is based
on an FTDI chip (FT\,245R).
The interface protocol allows channel-wise voltage setting, reading of
the currents and the over-current protection status. Ramping is
needed, since an immediate application of the operational voltage would
trigger the over-current protection due to the large load capacitance and
initial current of the G-APDs. A continuously increasing 3\,bit counter
is sent with every data transfer from the system controller to
allow for integrity checks. The cycle time for one command
is about 50\,$\mu$s. 
Over USB this speed is achievable only by using
bulk transfer of several commands.

\subsubsection{Interlock System, Remote Control and Heaters}
\label{sec:camera:aux:interlock}

Since summer 2012, the telescope is regularly operated remotely, i.e.\ without a
person present in the FACT counting hut. In order to allow this,
remote control is required and an interlock system was installed to
secure the unattended operation. The interlock system gets input
from the cooling unit (see section~\ref{sec:camera:aux:slow}), the fan
status of the G-APD bias system and remote commands transmitted via an
Arduino micro-controller. It controls the 48\,V main supply, the power
for the G-APD bias system and also the main AC lines for the drive
cabinet. Status information is provided locally by LEDs and by
the Arduino interface for remote access. The system operates as
follows:
\begin{enumerate}\setlength{\itemsep}{0pt} 
\item The 24\,V supply for the interlock system must be switched on, otherwise
  all other systems are blocked. This supply cannot be controlled
  remotely.
\item A button to overwrite the locked status has to be hold for some
  seconds, while the Pump-start button is pushed. This will turn on
  the pump of the cooling system and lead to correct
  readings of the flow and pressure status signal from the cooling
  unit. At this moment the overwrite button is no longer required and
  the system will self-hold the on-condition as long as pressure and
  flow are sufficient.
\item This enables the main camera supply as well as providing
  230\,V to the fan tray of the G-APD bias supply. The feedback from
  this fan-tray is used to enable the Agilent for the G-APD bias
  system as well as providing 230\,V to the G-APD bias crate.
\item Pressing the Stop button will interrupt the self-hold circuit and
  turn off the pump. At the same time all other devices are disabled and
  powered down.
\item In case of a cooling failure the system behaves in the same way
  as if the Stop button was pushed (see point 5).
\item The drive cabinet can be powered up and down independently by remote commanding.
\end{enumerate}
All above mentioned buttons can be activated remotely in the same way,
using the Arduino interface, including the read-back of the status
information. Moreover the interface allows the commanding of power
relays controlling the 3-phase main input lines to the drive cabinet.

The camera features a set of eight ohmic heaters, each providing 40\,W
at 24\,V. Two of them are installed per crate. They are controlled by
two thermo-switches, closing at 4\,\textdegree C and opening at
13\,\textdegree C, built into the camera. This system uses the 24\,V
from the interlock supply and thus is always enabled. Remote control
is extended to the AC power for all FACT computers using power sockets
with Ethernet interface.

\subsubsection{Cooling System}\label{sec:camera:aux:cooling}

The Hamamatsu MPPCs used as photosensors have a gain variation of
\(\sim\)5\%/K. They are glued onto the light concentrators, which are
in turn glued to the front window
(c.f.\ section~\ref{sec:camera:sensors}). It was desired to limit the
temperature variation across the sensor plane to 0.1\,K/cm and to a
maximum difference of 2\,K between any two sensors. A specific
operation temperature is not required. Since the electronics inside
the camera dissipates about 570\,W, an active cooling system is
required to prevent overheating of the components. In order to study
the thermal behavior of the camera, a thermal model of a quarter of
the camera was created by a specialized company (Thermotech division
of AMS Technologies), and the results were used to optimize the
design.

The camera is thermally divided into two compartments by means of a
baffle plate. The sensor compartment has no active cooling system. The
electronics compartment is water cooled. A central cooling plate
carries the four electronics crates and the supports for the DC-DC
converters. The water flows consecutively through the central plate and
the crates. Electronics boards are designed with heat spreading planes
and attached to the crates using CALMARK wedge-locks. The extra boards
(trigger master, slow control and fast signal distribution) are thermally connected to the outside of the
crates. The custom cooling unit is IP\,55 compliant. It is located at
the telescopes rotating platform inserted into a second casing
protecting it from splash waster. It provides a flow of 6.8\,l/min at
a maximum pressure of 5\,bar. A heat exchanger cools the water, containing an anti-freeze
(Glysantin\textsuperscript{\textregistered}
G48\textsuperscript{\textregistered}), to ambient air
temperature. About 10\,m long water hoses connect to the
camera. Switches reacting on a too low water flow or pressure deliver
an enable signal to the interlock system. Without a running cooling system the camera
is non-operable (see section~\ref{sec:camera:aux:interlock}).

\subsubsection{Slow Control Board}\label{sec:camera:aux:slow}

Inside the camera the temperature,
the humidity and all supply voltages are monitored continuously. A dedicated slow control board was
developed. It features an Atmel ATmega32L micro-controller and a Wiznet W\,5300
Ethernet interface for data transmission. A total of 148 channels can be multiplexed
onto a 24\,bit ADC, AD\,7719, with integrated current sourcing for temperature
probes. Concerning temperature and humidity, the slow control board reads out:
\begin{itemize}\setlength{\itemsep}{0pt} 
\item 31 temperatures close to the G-APDs in the sensor plane
\item 2 temperatures per readout crate, 8 in total
\item 2 temperatures per DC-DC converter board, 12 in total
\item 4 temperatures at the camera patch panel close to the cooling
  connections
\item 4 times the humidity, once close to each crate
\end{itemize}
As temperature sensors, PT\,1000 are used.

\subsubsection{Shutter}\label{sec:camera:aux:shutter}

The camera features a remote controllable shutter. Two linear
actuators of type LA\,23 (LINAK AG, Thalwil, Switzerland)
operate independently the two shutter
panels. The motors are controlled using an Arduino microcontroller
hosting a dual Pololu VNH-5019 motor driver board. The system controls
the power status of the motors, sets the motor speed and drives the
motors by providing power. It reads the motor currents and the
integrated hall sensors providing the motor positions. The system is
completed by an amplifier and filter card, for noise reduction on the
hall sensor reading and in order to obtain a second low current
measurement. An additional 24\,V low voltage supply is used to power
the system. Communication with the Arduino is established via Ethernet,
and a USB interface is available for debugging and firmware
upgrades. The system is operated independently of all other 
systems.
\begin{figure}[htb]
    \centering 
    \includegraphics[height=5.15cm]{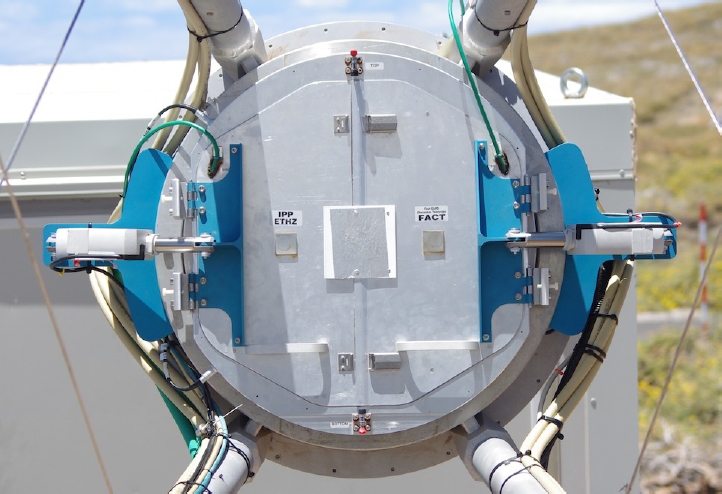} 
    \hfill
    \includegraphics[height=5.15cm]{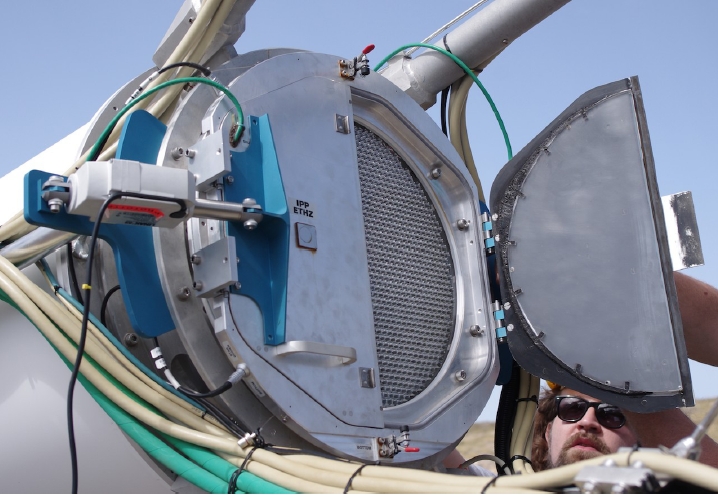}
    \caption{Photographs of the installed camera showing the
      shutter installation. Left: Closed shutter situation. The green
      cables visible on the top part of the picture provide power and
      control for the internal light pulser system. 
       The square plate in the center of
      the camera shutter is used as a screen, when pictures of star
      reflections are taken for the pointing calibration .
       Right: Situation with one
      half of the shutter opened. On the inner side of the opened
      half, the internal light pulser plate can be seen.}
    \label{fig:shutter}
\end{figure}

\newpage
\section{Online Software and Data Storage}\label{sec:software}

\subsection{Overview}

The slow control and data acquisition system is based on C++11
and implemented under recent Linux using several available standard
libraries. Its source code is open and stored in a svn
repository. As build system, GNU autoconf and GNU automake is
employed. The C++ boost libraries \cite{boost} are used as C++
extension.

The system is currently installed on the latest Long-term Support
version of Ubuntu (12.04\,LTS). All libraries in use, except CERN
developments, are available from their repositories. For global
authentication an LDAP server is in use. As database backend a
MySQL\,5 server is employed. Web-pages are served by an Apache\,2
web-server with php\,5. The whole system is composed by several
machines monitored by the Munin monitoring system \cite{munin}. The
system is reachable from the outside as a Virtual Private Network
(vpn).

\begin{figure}[tb]
    \centering
    \includegraphics[trim=1.5cm 1.17cm 1.15cm 1.5cm,clip=true,width=\textwidth]{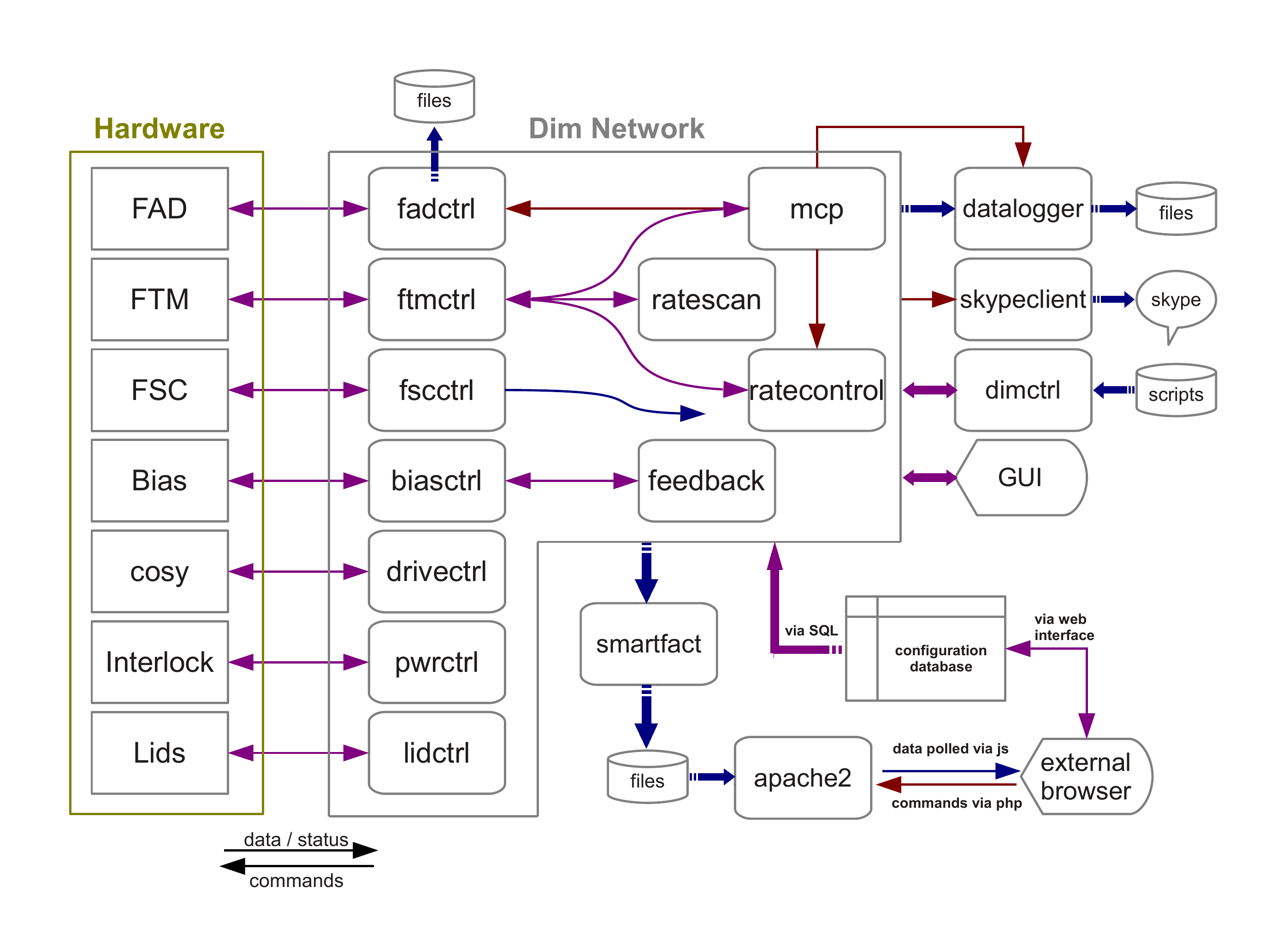}
    \caption{Schematics of the slow control system. Only the most
      relevant paths for data-exchange are shown. Magenta denotes
      command and data-paths, red the direction of commands and blue
      the direction of data-flow. The abbreviations denote the
      following components: FAD (digitizer boards), FTM
      (trigger master board), FSC (slow control board), Bias (bias
      voltage supply), cosy (drive system).  Other abbreviations are
      explained in the text.}
    \label{fig:slow-control}
\end{figure}

A general overview of the implementation is sketched in
figure~\ref{fig:slow-control}. The control system is split into several
programs, each responsible for a single task. This keeps algorithms
logically separated ensuring easy maintenance on the long
run. Each program internally implements an event driven state
machine. All communication is decoupled from the state machine
providing non-blocking behavior. Processing of all events is
synchronous ensuring robust operation and avoiding any possible race
conditions from the very first. Events can originate from the
inter-communication of programs, data received from the hardware or
the built-in console interface. Inter-communication is implemented
through CERN's Distributed Information Management System (DIM,
\cite{dim}). The hardware communication via Ethernet and USB is
managed by the asynchronous I/O library (ASIO). The readline-based
console interface features a history and tab-completion.

\subsection{Sub-components}

\subsubsection{System Configuration}

Each program retrieves its configuration automatically from different
sources, each with a different priority. The main sources are local
resource files and a configuration database. While the local files
feature mainly an easy way to alter the configuration for debugging
purposes, the database is used during operation. The database is
editable through a custom made php-based web-interface. Whenever the
database is altered, the history of changes is kept being able to
reconstruct the complete setup during any previous data-taking.
Configurations from two different times can easily be compared through
the web-interface.

\subsubsection{Inter-communication}

The communication between the programs implemented via DIM is
organized in a server-client setup. Every server is registered at a
central name-server allowing the access from any client by
name. Servers offer named commands and data services. After
subscription of a client to a data service, new data is distributed
automatically to all subscribed clients. Received commands and service
updates are both queued in an event queue of the state machine and
processed synchronously in the event loop.

The DIM network has been enhanced with a special service distributing
command and service descriptions of each program. These descriptions
allow to display help texts for commands and their arguments, and name
the elements of each service. Named elements allow a central data
collector to automatically subscribe to all services and write the
transmitted data to files. The files then contain a full description
of the data stored, and each change to the data services is
transparently logged.

\subsubsection{Data Storage}

All slow control data and logging output is collected by a central
{\it datalogger}. For each service, a file is kept open in FITS format
\cite{fits}. Every day at noon, a new file is opened. Since several
special features are needed in this context, like updating the headers
during streaming, the FITS output is based on the cfitsio-library and
its C++ wrapper CCfits \cite{ccfits}. Files in FITS format contain a
complete description of the data stored. This description is
automatically constituted from the format description available for
each DIM services and the custom element description offered by each
server. The FITS format was chosen because it features
self-contained data by storing a full format description in the
header. It is easy to implement, widely spread in astronomy and
several tools and libraries are available.

The raw data received from the data acquisition hardware is
preprocessed by the event-builder, both described in more details in
section~\ref{sec:daq}, and also stored in FITS format. The simplicity of
the format allows sustained streaming of the data implemented in a
custom FITS streamer class. A complementary class for reading the data
as a continuous stream is also available. Advantages of a custom
implementation are the absence of any non-essential feature slowing
down the streaming process and the possibility to compress and
decompress data streams during reading and writing on the fly using
the zlib-library~\cite{zlib}. The FITS CHECKSUM \cite{checksum}, a
simple integrity check, is calculated and checked on the fly whenever
a file is written or read.

\subsubsection{Dimctrl}

As a central DIM client, a program called {\it dimctrl} is available.
Although each program can be accessed individually from its console
interface, {\it dimctrl} features a central command interface.
Loading batch scripts from file allows to automate simple processes.
In addition, a JavaScript (ECMA script) interpreter is available,
implemented by means of Google's JavaScript engine V8 \cite{v8}, which
allows full access to the DIM network and access to the
database. Running JavaScripts decoupled from the DIM network in a
sandbox allows to stop them at any time. This allows to recover a
hanging system in case of an unforeseen event without the need to
terminate {\it dimctrl}. By this, no log-in to the local machines is
required anymore. Database access allows to directly retrieve and
execute the observation schedule, provided by an automatic scheduling
program or interactively by the user.

\subsubsection{Graphical User Interface}

The graphical user interface (GUI) is implemented using Qt\,4 and the
qt-designer \cite{qt}. CERN's root \cite{root} is used to display
histograms and graphs, and a two-dimensional camera display is drawn
through an OpenGL \cite{opengl} context. It is the only program
implementing its own Qt\,4-based event loop not offering itself as
server, thus not being able to receive commands from any client. The
GUI subscribes to all DIM services of interest and posts their updates
to the Qt\,4 event loop from where the corresponding GUI elements are
altered accordingly. User events are directly emitted as DIM commands
to the corresponding servers.

The advantage is the complete decoupling of the graphical interface
from the control system, thus the decoupling of a part known to be a
primary source of problems during data-taking. The drawback of the GUI
connected to the DIM network is the missing tolerance of the DIM
network against slow network connections or blocking handlers used to
process the incoming data. Consequently, a slow network connection and
the transmission of large data as service update or command can block
all DIM clients at once. This makes the GUI only an option for local
control or to be exported indirectly, e.g.\ via a vnc connection. A
central DIM engine could buffer all service updates and selectively
distribute those data to other clients, taking into account connection
speed. However, this would require additional development work.

\newpage
\subsubsection{Smartfact} 

\begin{wrapfigure}[30]{r}{.41\textwidth}
  \centering
  \includegraphics[width=.41\textwidth]{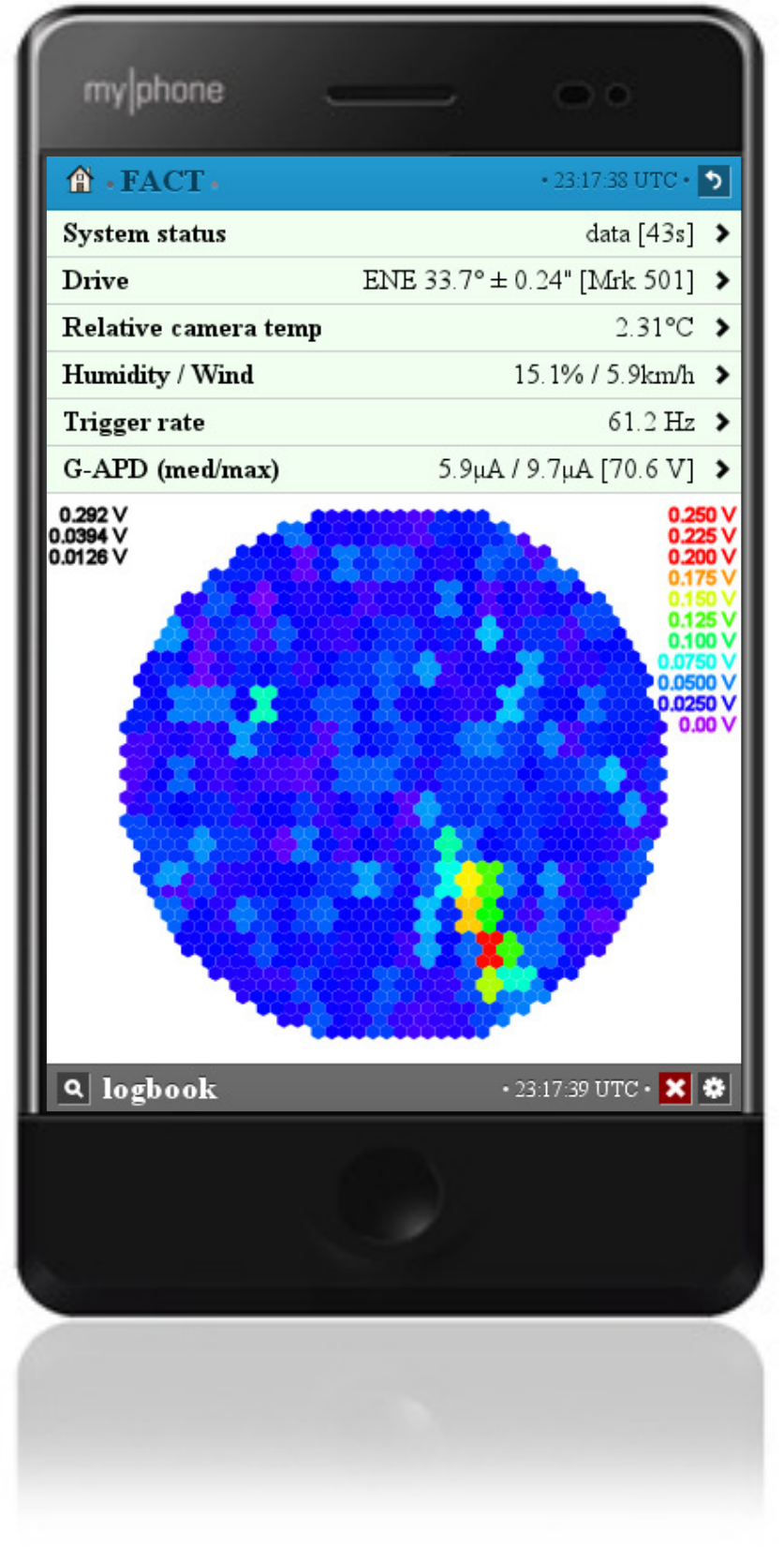} 
  \caption{Screenshot of the Smartfact web-interface during
    data-taking.  To maintain a low network traffic only an average
    value per bias-voltage channel is transmitted and displayed
    instead of one value per pixel.}
  \label{fig:smartfact}
\end{wrapfigure}

For a stable remote operation of the telescope, a complete decoupling
of the interactive user-interface from the control system is
necessary. This is done keeping all communication within the DIM
network locally and implementing external communicating with the user
through a web-server.

The Smartfact user-interface, as shown in figure~\ref{fig:smartfact}, is
based on a small JavaScript. The JavaScript is once retrieved from the
web server. To load a page, the JavaScript reloads a small description
file from the server. From this description, the displayed HTML is
constructed dynamically. The displayed data is reloaded every few
seconds from the server. Both, structure-files and data-files, contain
only the relevant information in a compressed format to keep the
network traffic below 1\,kB/s. This allows to control the system even
over extremely slow connections. The displayed graphics is rendered
using canvas-elements defined in HTML\,5.

The back-end of the Smartfact interface is the {\it smartfact} program
subscribing to all services of interest. Either triggered by the
reception of a service update or in regular intervals, the data-files
on disk are updated. For simple astrometric calculations, e.g.\ an
overview of the zenith angle of all possible sources over the whole
night, libnova \cite{nova} is used. In addition, the user can emit
commands to the system. This is implemented by a php-script accessed
from the web interface. The php-script then triggers start or stop of
a JavaScript in {\it dimctrl}. Form elements defined in the
structure-files allow to customize the executed action by the user
sending additional data to {\it dimctrl}.

\subsubsection{Automation/Scheduling}
Using the JavaScript interface with data-base access, automatic
operation based on a pre-defined scheduling is currently under
development and being tested. In a first step, the nightly schedule
will be retrieved from a database and operations will take place
accordingly. To fill the database in advance, a graphical
web-interface is available, which allow to easily adapt the schedule
on a nightly bases and acts as an archive for past schedules. For each
night, plots are available which simplify scheduling, such as the
zenith angle of each source versus time. Sources and times can be
entered manually or be changed by checkboxes on a list of visible
sources and sliders displayed in the plots.

\subsubsection{Electronic Logbook}

As a shift logbook the MyBB forum software \cite{mybb} is
installed. Each nightly shift is realized as a single thread. The
advantage is that no special maintenance is necessary and standard
software can be used. After or even during the shift other
collaboration members can directly comment on logbook entries
simplifying the later interpretation for the person in charge of data
analysis. With some installed extensions, MyBB directly offers an
overview calendar, easy editing (markdown language), a version for
mobile devices, user authentication through the global LDAP password
and marking entries as different priority. A possibility to add images
or other documents and a search function is provided. Individual
access rights can be customized to a large extend. The personal
authentication system allows to keep easily track of the origin of
each entry. Other forum categories offer the possibility to store
related information within the same framework.

\subsubsection{Individual Programs}

Individual hardware components are controlled by eight programs. The
40 digitizer boards, the trigger master board and the slow-control
board are controlled by {\it fadctrl}, {\it ftmctrl} and {\it
  fscctrl}, respectively, over custom Ethernet protocols. The bias
voltage supply is controlled by the {\it biasctrl} over USB. Two
Arduino-based boards controlling the power switches and the lids are
operated via a simple built-in web-server from {\it pwrctrl} and {\it
  lidctrl}. The received HTML is parsed using Qt\,4's HTML parser. To
be able to re-use the MAGIC control software for the drive system, the
{\it drivectrl} mimics the MAGIC slow control communication. The
temperature dependence of the G-APD response is compensated by the
{\it feedback}, using as input the readings from 31 temperature
sensors in the sensor compartment. These values are distributed by the
{\it fscctrl}. Taking the current readings for all 320 bias voltage
channels, it also corrects the applied voltage for the corresponding
voltage drop in the system. At the beginning of the data-taking the
internal resistance of each channel of the bias power supply is
calibrated. A {\it ratecontrol} adapts the over-all trigger threshold
at the beginning of each run to the current light conditions. During
the run, outliers induced by, e.g., reflections of bright stars are
treated individually to keep the data rate reasonably low. The Master
Control Program ({\it mcp}) coordinates all actions necessary to
configure and start runs of different run-types. Ratescans measuring
the dependence of the camera trigger rate from the trigger threshold
are controlled by a {\it ratescan} program.

Additional auxiliary information is received from the weather station
of the TNG telescope and the weather station of the MAGIC telescope by
accessing their web-pages with {\it tngweather} and {\it
  magicweather}, respectively. The TNG web-page features XML data
which is parsed using the Soprano library \cite{soprano}.

A special server is the {\it skypeclient} which communicates over
Skype's DBus interface \cite{skype} with an open skype client. This
allows the control of a skype session remotely and allows sending of
alarm SMS in case of important alerts. Even ringing a phone is
possible.

\subsubsection{Feedback}

Several ohmic resistors in series (shunt, output protection, cable, filter and bias resistors)
introduce a voltage drop between the bias voltage supplies and
the G-APDs depending on the current flow. The
drawn current varies with the level of the diffuse night-sky
background light in all channels, and due to bright stars in
individual channels. Since most G-APD properties depend on the applied
voltage, a voltage correction is necessary. The feedback resistor of the HV
operational amplifier of each bias voltage channel is
calibrated by measuring the current for a set-point 3\,V below the
G-APDs breakdown voltage, thus no more operating the diode in Geiger-mode.
Correcting the measured currents for the current loss at
the parallel resistance, the voltage drop at each G-APD is calculated.
The feedback
program increases the applied voltage accordingly until the voltage
calculated at the G-APD reaches the desired set-point. Included in
this procedure is a linear correction of the gain
changes induced by the temperature dependence of the G-APD.
As reference, the average measured temperature in the
sensor compartment is used.

\subsubsection{Ratecontrol}

The trigger system measures the trigger rate of each of the 160
comparator outputs (patch rates) and the trigger rates at the output
of the 40 N-out-of-4 logic (board rates). Since the N-out-of-4 circuit
is also sensitive to the width of the input signal, electronic-noise
induced triggers are filtered out and therefore patch and board rates
are not exactly correlated.

At the beginning of each run, the median trigger rate of all board
rates is regulated such that the accidental trigger rate induced by
NSB is well below the trigger rate
induced by charged particles. This step intentionally ignores single
patches with higher rates as an effect of bright stars. The resulting
threshold is kept constant throughout one run (usually 5\,min) to
ensure a stable energy threshold during each run. During the run,
individual boards with exceptionally high trigger rates are identified
and the threshold is increased for the patch with the highest trigger
rate within that board. For every patch with an increased threshold
and a trigger rate well below the limit for regulation, the threshold
is decreased.

\subsection{Data Acquisition}\label{sec:daq}

\begin{figure}[t]
  \centering
    \includegraphics[trim=0.1cm 11.0cm 4.0cm 0.5cm,clip=true,width=0.85\textwidth]{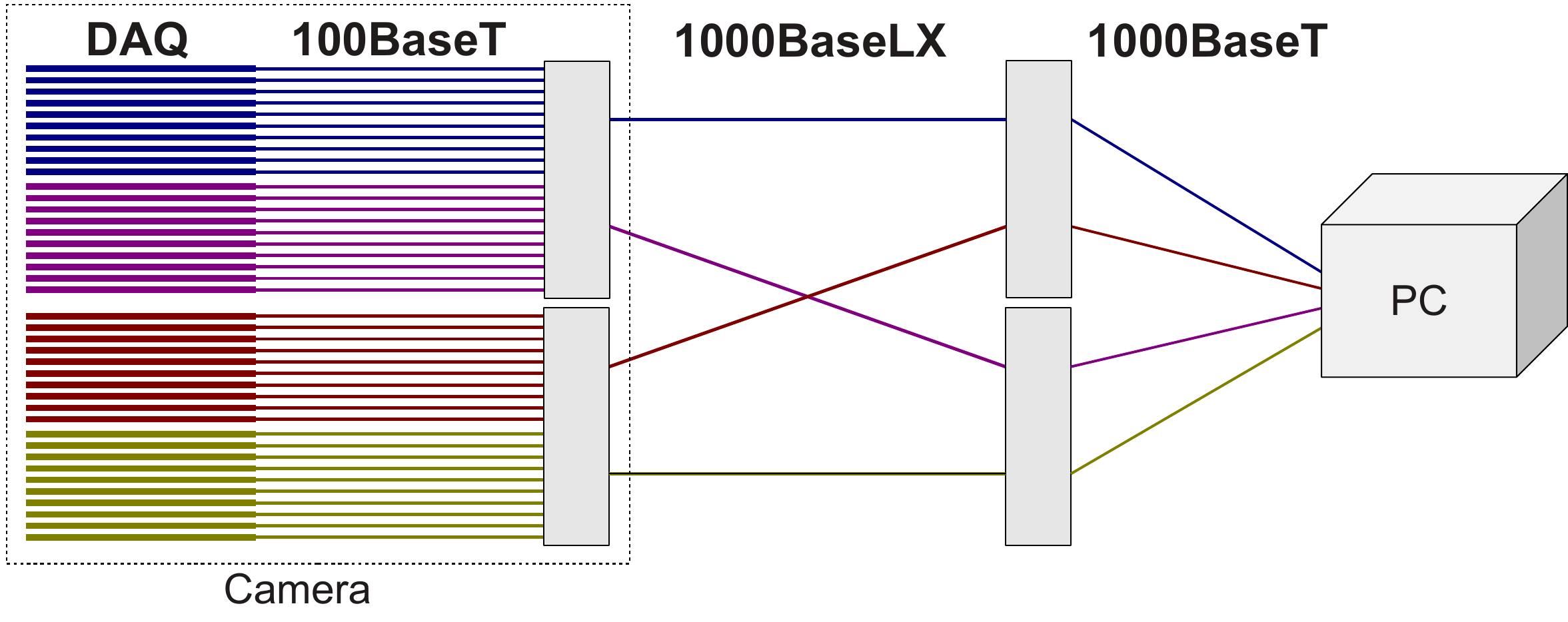}
    \caption{Overview of the FACT data acquisition system. From left
      to right four crates (10 boards each), the switches inside the
      camera, the switches in the counting house and the data
      acquisition computer are indicated. On the top, the Ethernet
      connection standard is given.}
    \label{fig:ethernet_DAQ}
\end{figure}

As described in section~\ref{sec:camera:electronics}, the camera contains
40 digitizer boards distributed over four crates. Each board
reads out 36 pixels and is equipped with a commercial 100\,MBit
Ethernet interface delivering a maximum sustained throughput of
7\,MB/s. Inside the camera there are two commercial low-power smart
Ethernet switches (D-Link DGS\,1224T), both having two 1000\,Mbit fiber
and 22 100/1000\,Mbit copper connections. The switches are configured such
that each crate has a dedicated fiber assigned. This ensures that the
maximum sustained data rate deliverable by all boards does not
overload the bandwidth of one fiber. The two other boards with an
Ethernet interface, namely the trigger master and slow control board,
are connected to the remaining copper connectors. For all Ethernet
connections between the boards and the switches, slim cables of type
Wirewin Cat.6 UTP are used. The data transmission from these switches
to the backplane of the camera, and from there to the counting house,
is done using optical fibers (Huber + Suhner, Masterline). In the
counting house the fibers are connected to two more Ethernet switches
of the same type having all computers needed to communicate with the
camera connected to them. Even in case one of the two fiber pairs coming
from the camera would fail, the system could be operated (at reduced performance).
In figure~\ref{fig:ethernet_DAQ} an overview
of the architecture is shown. In addition, all computers in
the counting room are connected to an internal network for data
exchange.

The Data Acquisition (DAQ) computer (and its spare equivalent) has
four dedicated 1000\,Mbit Ethernet ports to communicate with the
camera. All switches, as well as the ports of the computer, are
configured to provide a bandwidth of 1000\,Mbit between the computer
and the camera, individually for all four crates. This allows to reach
the maximum transfer rate of 7\,MB/s to each of the 40 digitizer
boards. Tests have shown that a sustained bandwidth
between the camera and the computer exceeding 250\,MB/s can indeed be
maintained for extended time periods. When all 1024 samples per channel
are read out, this bandwidth allows a sustained data rate
corresponding to an average trigger rate of 70\,Hz. When only 300
samples are retrieved, a trigger rate of 230\,Hz can be achieved. This
is well above the expected physics trigger rate for a telescope with a
9.5\,m$^2$ mirror area. Ethernet interfaces and components with a ten
times higher performance could have been integrated, but they
would have had a significantly higher power consumption and would not
have been necessary. Therefore these ideas were discarded.

Each of the 40 digitizer boards sends its data asynchronously
to the computer, adding a header with additional information as a
continuous event number. Based on this header information, a multi-threaded
event builder program collects the data for one event from the different
boards and writes the complete events to disk. In order to reduce in the future
the amount of data written to disk, the event builder has
already been prepared to apply a software trigger. Studies on an optimal trigger algorithm
are ongoing.


\subsection{Data Taking Procedure}\label{sec:data_taking}

Data taking is currently automated using batch scripts of commands
executed by {\it dimctrl}. In this way, data integrity and data
consistency can be ensured. Recently, scripts written in JavaScript
are being implemented which allow a more complete automation like automatic
error handling and retrieval of the schedule from an external source.

After power-on of all components, several checks are performed
ensuring that all components got properly initialized. In particular,
this is necessary for all components equipped with the Wiznet Ethernet
chip, for which, due to a bug in the chip, sometimes negotiating of
the connection speed with the connecting switch fails. This would
result in unstable network connections. Subsequent to a proper
initialization, a first DRS calibration is performed. For this, runs
are taken with pseudo-random triggers with the G-APD bias voltage
switched off. They are used to calibrate offset and gain of each
capacitor as well as a special offset introduced during readout into
each logical sample. A special run digitizing a square signal is used
for calibrating the timing behavior of each sample.
Data taking is then started with the pseudo-random trigger and bias voltage on,
while keeping the lids closed. This way the G-APD properties can be cross-checked
by extracting their single-pe spectrum \cite{feedback}.

Observation of sources is currently carried out in two repeated
blocks, each consisting of four runs with a duration of five
minutes. For each observation, a dedicated DRS calibration is taken to
account for temperature changes during the night. Each block is preceded by
a pseudo-ranomly triggered pedestal run and a light-pulser run to have
a cross-check reference for baseline and gain calibration. Each run is
interleaved with pedestal and light-pulser events at a rate of 1\,Hz.
Currently, 150\,ns sampled with 2\,GHz are recorded in each
event. During each night, at least one ratescan is recorded for a
continues monitoring of the atmospheric conditions as well as the
performance of the system. Ratescans and other special technical
measurements are also automated by batch scripts.

Offline tools to easily access the data in the FITS files from the console and display the raw
data as well as summary data (e.g.\ maximum value in the pipeline) are
available.

\subsection{Data Center}\label{sec:data_center}

After being recorded, the raw data are compressed using gzip to save
disk space. Afterwards, files are automatically copied to the data
center at ISDC and backed up from there to the data center in
W\"urzburg. Already on-site at La Palma, quality as well as
consistency checks are carried out. Information on the data and their
quality are extracted from both, the raw and auxiliary data, and
inserted to a MySQL database, from where they can be queried via a web
interface. The stored values are later on used in the data
selection for the physics analysis. At ISDC, again data consistency
checks are performed before the validated data files are inserted into
an archive. From this archive all data can be downloaded in FITS
format. Further automatic processing steps have already been
implemented, like calibration, signal extraction, image cleaning and
calculation of image parameters. Also the Monte Carlo production and
processing is already automatized.

\newpage
\section{Conclusions}

During the past years, the first Cherenkov telescope using solid-state
photosensors, Geiger-mode Avalanche Photodiodes, has been designed,
built and set up by the FACT collaboration (see
also~\cite{FACT,ICATPP,IEEE}). It is in operation since Oct.\ 2011,
c.f.\ \cite{Cern}.

As telescope structure, the mount of the previous HEGRA array is in use. Its 32
old mirrors have been replaced by refurbished mirrors with an about 10\%
larger reflective surface, 9.51\,m\(^2\) in total. The mirrors were
measured to have a peak reflectivity of about 90\%. The drive
system has been replaced with a state-of-the-art drive system as
applied on both MAGIC telescopes. The focal distance is 4.9\,m
yielding \(F/D\approx1.4\).

All necessary electronics, except the power supplies, are integrated in
the camera body resulting in a total camera weight of about 150\,kg. A
passive water-cooling is applied. The total power consumption of the
camera electronics is less than 0.5\,kW.

The camera has a total disc-like field-of-view of 4.5\textdegree{}
composed of 1440 pixels \'{a} 0.11\textdegree{}. Each pixel is
equipped with a G-APD and a solid light concentrator (cone), made of
plexiglass by injection molding, and glued to a protective front
window. The grid distance of the hexagonal pixels is 9.5\,mm. While
the sensitive area of each G-APD is a square of 9\,mm\(^2\), the cone
exit area is 7.8\,mm\(^2\). The cones were measured to have a peak
transmission of 98\% and the G-APDs have a peak photon detection
efficiency (PDE) of \(\sim\)33\%.

The signal of nine pixels is summed to form a single trigger patch.
Patches are arranged on an ideal hexagonal grid (see
\cite{twepp12}). Before discrimination, clipping (shortening of the
signal in time) is applied. As a noise filter, an additional
N-out-of-4 logic is applied on the sum of four of the sum-signals. The
trigger signal is distributed to the readout electronics, which are
based on the DRS\,4 ring-sampling chip allowing a time jitter between
different channels as low as about 300\,ps (depending on the sampling frequency).
Each trigger patch is sub-divided into a compact four-pixel
and five-pixel group. The G-APDs in one of the sub-groups share the same
bias voltage, so in total 320 bias voltage channels are used.

\subsection{Problems during Construction, Final Tests, and Operation}

The industrial production of the light concentrators
turned out to be more challenging than anticipated. Together with the producer the quality had to
be improved gradually, and several selection measurements were necessary. Also the glueing
of the concentrators to the G-APDs required some test runs and manual work. However, it should
be noted that the primary goal was not to study or develop automated mass production techniques,
but to equip the 1440 pixels of the FACT camera with light concentrators in a reasonable time.  

After assembly of the camera, three pixels have been found
dead thus delivering no signal at all. Since access to the sensor
compartment is very difficult due to limited space, it was decided
that the corresponding 2\textperthousand{} loss is acceptable. Another three pixels
where identified to have a too high bias-voltage, most probably due to
a wrong serial resistor. They are switched off for the trigger during
normal operation. Like for the dead pixels, it was decided that repair works would
take too long and compromise the project start before winter.
Furthermore, three patches
with one signal short-cut between two neighboring pixels were identified.
Since these twin-pixels deliver an average signal, no
influence on the trigger could be found. For
data-analysis they act like a locally slightly larger pixel. Due to
the high granularity of the image, no strong influence could be found.

During 1.5 years of operations, five hardware problems
happened. Once, the firmware of the bias power supply got lost and had
to be reprogrammed. A reason could not be identified. In summer 2012,
the fuse of a DC-DC converter board in the camera blew. Since no
obvious reason could be found, the origin might have been the fuse
itself. This was fixed a few days later by an access to the camera. In
winter 2012/2013, one bias voltage channel broke. Replacing the board
in the power supply fixed the problem. Furthermore, a few times
the clock conditioner on the trigger master board did not lock correctly
after the startup of the camera electronics. To avoid lengthy multiple
reboots, it was decided to keep the camera electronics powered all the time
(in standby). The fifth problem was a bad cable connection inside the cabinet of
the drive system (in the counting room). It should be emphasized that all of the
above-mentioned problems are not related to the G-APDs.



\subsection{Achievements}

For the first time, an IACT not using classical
PMTs, but G-APDs, for photon detection has been constructed and put into operation. The applied photo sensors were
ordered off the shelf and have proven to be all within their
specifications. The precision of each device and the measurement of
the breakdown-voltage provided by the manufacturer was good enough
that they could easily be sorted and a single voltage can be applied
to several sensors at once.
Keeping the overvoltage constant and homogenous throughout the camera during data taking could be achieved by a simple voltage adjustment, based on the consumed current and the ambient temperature.
Details are discussed in a dedicated
performance paper \cite{feedback}.

\begin{figure}[htb]
  \centering
    \includegraphics[trim=0.5cm 0.0cm 3.3cm 2.5cm,clip=true,width=0.49\textwidth]{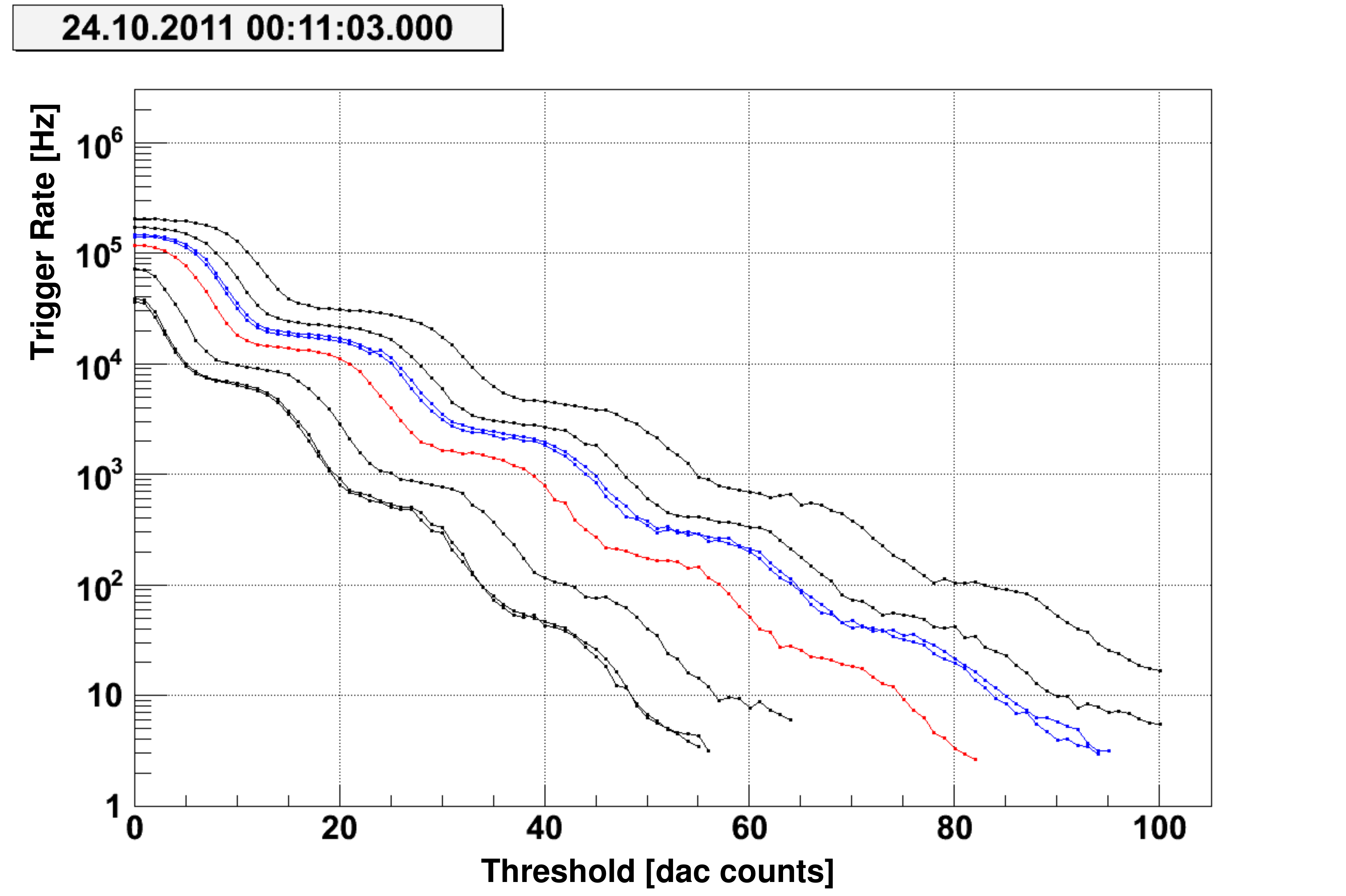}
    \includegraphics[trim=0.0cm 0.0cm 2.0cm 1.2cm,clip=true,width=0.49\textwidth]{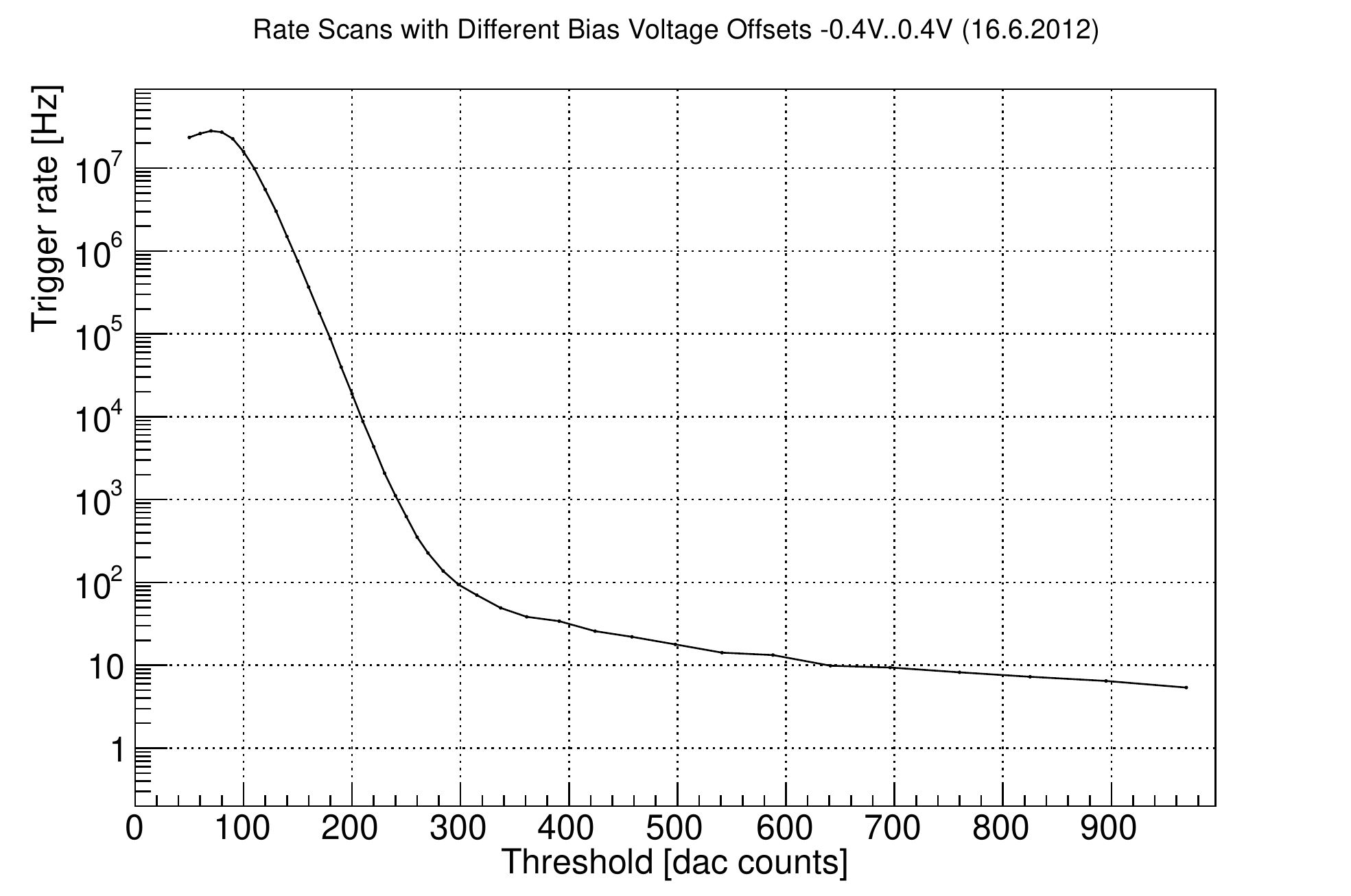}
    \caption{Left: Trigger rate in Hz versus comparator threshold in
      dac-counts with shutter closed.
      For this measurement, all pixels
      except one have been switched off in the trigger
      branch. Different curves were taken with different bias voltages
      applied. In all cases, the single-p.e.\ response is
      visible. Right: Overall camera trigger rate in Hz versus
      comparator threshold in dac-counts taken during
      dark night and good weather conditions. The trigger rates below a
      threshold of about one hundred dac-counts are influenced by
      saturation effects. Up to about 250 dac-counts, triggers of
      coincident photons from the diffuse night-sky background
      dominate. Above a threshold of about 300 dac-counts, triggers
      from hadron-induced air-showers dominate. The slope of the
      hadron branch exhibits a good correlation with the slope
      expected from the proton spectrum.}
    \label{fig:ratescan}
\end{figure}

The readout system has single-p.e.\ resolution. More details on that can be found in
\cite{feedback}. A timing resolution of about 600\,ps has been achieved for signals in the
order of 5\,p.e., including contributions from the whole signal chain (e.g.\ from the mirrors
and the DRS chips). The precision of the trigger chain
is comparable to the single-p.e.\ resolution of the readout chain,
visible when measuring the trigger rate of a single pixel (all pixels
except one switched off in a trigger patch) versus the comparator
threshold, as shown in figure~\ref{fig:ratescan} (left).

\begin{figure}[hbt]
    \centering
    \includegraphics[width=.49\textwidth]{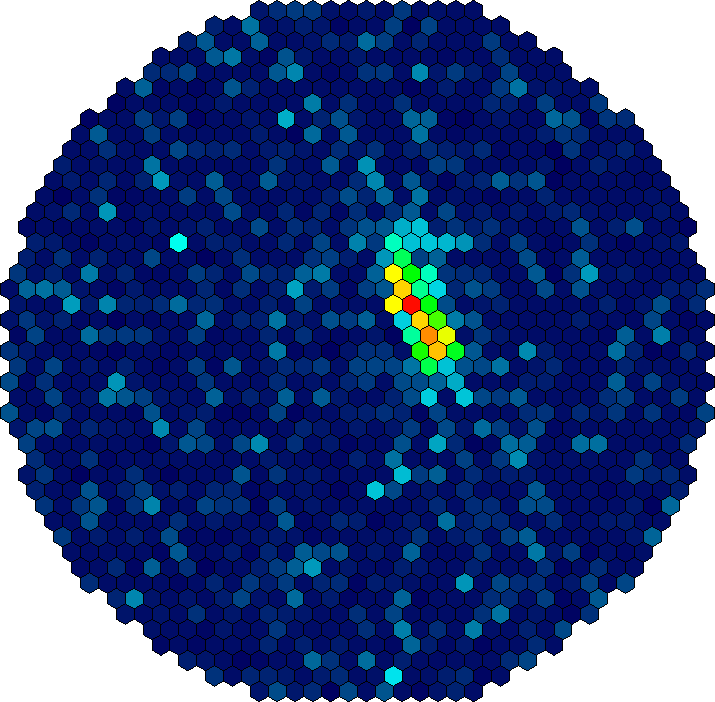}
    \includegraphics[width=.49\textwidth]{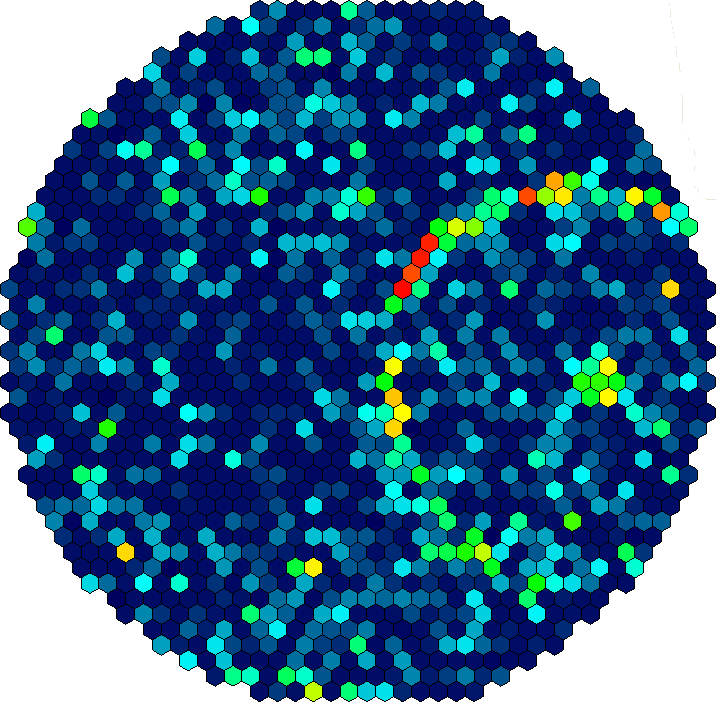}\\[2pt]
    \includegraphics[width=.49\textwidth]{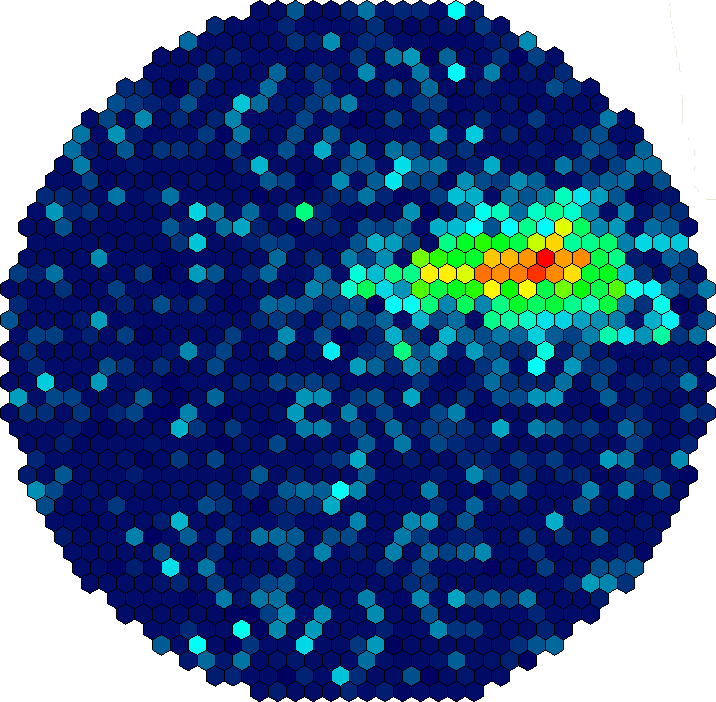}
    \includegraphics[width=.49\textwidth]{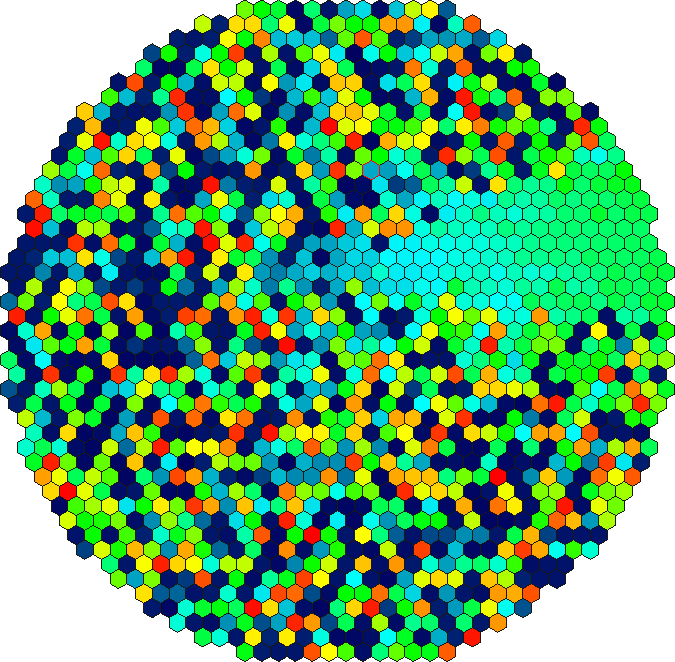}
    \caption{Examples of air shower images recorded by FACT. The two images
    at the top correspond to a gamma candidate and a muon event, with the color code
    representing the pixel amplitude. At the bottom a proton shower is shown, both in amplitude (left)
    and signal time (right).}
    \label{fig:showers}
\end{figure}

Measurements of the total trigger rate versus comparator threshold,
taken during dark nights (see figure~\ref{fig:ratescan}, right) as well
as moonlit nights, show the expected behavior and prove that the
trigger works well. For a reasonable threshold, a rate in the order of
50\,Hz\,--\,60\,Hz is achieved during data taking. Some example events are shown
in figure~\ref{fig:showers}. In order to
adapt the trigger thresholds to different light conditions, a
rate-control has been implemented keeping the camera response
homogeneous and stable through single data-taking runs.

Keeping the sensor and trigger response stable and
homogeneous, first physics results have been obtained and published in
\cite{Gamma2012}. Except for the above-mentioned minor problems, the
operation of the telescope is stable since the moment it was put into
operation. First air shower events were recorded immediately after the
installation of the camera \cite{Cern}.

The consistent concept of the slow-control software features a robust
and stable operation software-wise. A powerful local graphical
interface is available for telescope and camera monitoring and
control. A web-based interface with low bandwidth requirements has
been implemented for remote operation. Telescope control is available
through a JavaScript interpreter and fully automatic operation is
already in the test-phase. A first automatic data-analysis running on
site is available and can soon be used to deliver triggers for
target-of-opportunity observations to other observatories. Data is
written in a self-explanatory format (FITS), with the goal to make it
publicly available.

\subsection{Possible Future Improvements and Outlook}

Although the telescope is working very reliably,
still some improvements can be
implemented. On the hardware side, this is mainly an alignment system
for the mirrors, which is currently in the design phase. On the software side,
this is the implementation of a software trigger to suppress
background events triggered from night-sky background.
Another possibility is
to upgrade the system such that the voltage applied to the 
G-APDs is adjusted without involving a computer in the control
loop. However, since this equires not only a new firmware for the controller in the
supply crate, but also a re-design of the hardware, it will probably not happen in the near
future.

A nearly complete remote operation of the telescope has already been
achieved. Currently under development is the automated
operation. First tests of the system have already been performed and
automatic operation will start within the coming months. Of
course, robotic, i.e.\ self-sustained, operations are limited by a
hardware not primarily build for this purpose, so that the need for a
remote shifter during night is still expected.

Geiger-mode avalanche photodiodes (G-APD) have turned into a great
option for Cherenkov telescopes, in particular also for the CTA project \cite{CTA}. They
have kept their promise to be easy to handle and to allow for a very stable
operation also during moontime conditions. Even though production of
solid light concentrators, and their glueing process, needs further
investigation to be considered for mass production, the FACT concept
has turned out to be a benchmark for future Cherenkov
telescopes. Considering the fast drop in price for solid state devices
(today's price for G-APDs is already about a factor of ten lower than
when the devices for FACT were bought), the construction of the FACT
camera has proven that monitoring of the brightest sources with
single Cherenkov telescopes becomes affordable and that 24/7
monitoring on long-terms is possible.

Source monitoring with FACT has already been on-going since the installation
of the camera, thus during the commissioning of the instrument.
In the future the priority will be the stable and
uninterrupted operation of the system to ensure continuous long-term
monitoring of the brightest known blazars \cite{bretz08} and to act as a flare-trigger
for other telescopes.

\acknowledgments
The important contributions from ETH Zurich grants ETH-10.08-2 and
ETH-27.12-1 as well as the funding by the German BMBF (Verbundforschung
Astro- und Astroteilchenphysik) are gratefully acknowledged. We thank the
Instituto de Astrofisica de Canarias allowing us to operate the telescope at
the Observatorio Roque de los Muchachos in La Palma, and the
Max-Planck-Institut f\"ur Physik for providing us with the mount of the
former HEGRA CT\,3 telescope. We also thank the group of Marinella Tose 
from the College of Engineering and Technology at Western Mindanao
State University, Philippines, for providing us with the scheduling
web-interface.


\newcommand{\link}[1]{{\href{#1}{\url{#1}}}}

\end{document}